\documentclass{elsart}
\usepackage{graphics}

\begin{document}

\def\etal{{\it{}et~al.}}        

\def\psfigure#1#2{\resizebox{#2}{!}{\includegraphics{#1}}}

\textfloatsep=1cm               
\floatsep=1.5cm                 

\date{February 12, 1997}
\journal{the Journal of Theoretical Biology}

\begin{frontmatter}
\title{A model of mass extinction}
\author{M. E. J. Newman}
\address{Cornell Theory Center, Rhodes Hall, Ithaca, NY 14853.}
\address{and}
\address{Santa Fe Institute, 1399 Hyde Park Road, Santa Fe, NM 87501.}
\begin{abstract}
  A number of authors have in recent years proposed that the processes of
  macroevolution may give rise to self-organized critical phenomena which
  could have a significant effect on the dynamics of ecosystems.  In
  particular it has been suggested that mass extinction may arise through a
  purely biotic mechanism as the result of so-called coevolutionary
  avalanches.  In this paper we first explore the empirical evidence which
  has been put forward in favor of this conclusion.  The data center
  principally around the existence of power-law functional forms in the
  distribution of the sizes of extinction events and other quantities.  We
  then propose a new mathematical model of mass extinction which does not
  rely on coevolutionary effects and in which extinction is caused entirely
  by the action of environmental stresses on species.  In combination with
  a simple model of species adaptation we show that this process can
  account for all the observed data without the need to invoke coevolution
  and critical processes.  The model also makes some independent
  predictions, such as the existence of ``aftershock'' extinctions in the
  aftermath of large mass extinction events, which should in theory be
  testable against the fossil record.
\end{abstract}
\end{frontmatter}

\section{Introduction}
\label{intro}
Building on ideas first put forward by Kauffman~(1992), there has in recent
years been increasing interest in the possibility that evolution may be a
self-organized critical phenomenon~(Sol\'e and Bascompte~1996, Bak and
Paczuski~1996).  The basic argument is that the species in an ecosystem are
not independent of one another, but interact, and that these interactions,
in combination with the spontaneous mutation and genetic variation which is
always present in populations, can give rise to large evolutionary
disturbances, termed ``coevolutionary avalanches''.  In this paper we
investigate some of the evidence which has been put forward in favor of
these processes, and ask whether these data really demonstrate what it is
claimed they do.  First, however, we give a brief summary of the
fundamental concepts involved---coevolution and self-organized
criticality---and review some of the theoretical work which has been done
in the area.

Coevolution arises as a result of interactions between different species.
The most common such interactions are predation, parasitism, competition
for resources, and symbiosis.  As a result of interactions such as these,
the evolutionary adaptation of one species can force the adaptation of
another.  Many examples are familiar to us, especially ones involving
predatory or parasitic interactions.  The evolutionary pressure which the
cheetah and the antelope place on one another to run faster is a case in
point, or the perpetual evolutionary arms races which take place between a
disease and its host.  The pressure on the trees of a forest canopy to grow
ever taller is an example of coevolution because of competition, in this
case for sunlight.

If the evolution of one species can provoke an evolutionary change in
another with which it interacts, then presumably it is possible for the
change in that second species to provoke one in a third, and so on.  Such
an evolutionary chain reaction has been dubbed a coevolutionary avalanche,
and if such avalanches really existed, they would raise some interesting
questions.  It has been suggested for example that they could provide an
explanation for the observed high rate of species extinction in the fossil
record (Bak and Paczuski~1996).  It is known that almost all of the species
which have lived on the Earth are now extinct (Raup~1986).  Only about one
in a thousand of those which have ever existed are alive today, and most of
the others didn't last very long---less than ten million years in most
cases.  Some of these were wiped out by well-documented cataclysmic events.
The K--T boundary event is the most famous example, caused perhaps by the
impact of a meteor (Alvarez~\etal~1980, Sharpton~\etal~1992, Glen~1994).
However, the majority of extinctions have no known cause.  It is possible
that some of them were the result not of environmental disasters but simply
of natural evolutionary processes.  If a coevolutionary avalanche of
sufficient size were to pass through the ecosystem, causing the evolution
of thousands of species to new forms, it is conceivable that certain
species would find their livelihoods destroyed by the changes, and be
driven to extinction.

An alternative and subtly different scenario is that of large-scale
``pseudoextinction'', the competitive replacement of species by their own
descendents.  It is the central tenet of the theory of evolution that
organisms undergo mutations which in rare cases make them better able to
survive and reproduce, with the result that the descendents of the mutant
supplant the ancestral species, which usually becomes extinct.  In the
fossil record this pseudoextinction process is discernible from true
extinction---the death of a species without issue---and traditionally true
extinction has provoked more interest, since the processes by which it
takes place are largely a mystery.  However, if the coevolutionary
avalanches described above really do occur, then presumably they give rise
to wholesale pseudoextinction, and this could have a significant effect on
the rates of species turnover.  In theory, one might look for evidence of
these sweeping waves of pseudoextinction in the fossil record, though no
such quantitative study has been done, and it is not even clear that the
available data are equal to the task.

In any case, whatever the particular extinction process we are interested
in, we are now led to another question.  If coevolutionary avalanches are
to produce an effect large enough to be seen in the fossil record, then the
avalanches must be very large---of a size comparable with the size of the
entire ecosystem.\footnote{Alternatively there could be very many small
  avalanches.  However, the self-organized critical theories focus on the
  large avalanche possibility.} It seems not unreasonable to hypothesize
however that the typical avalanche would affect only a handful of species.
The intriguing theory which has been proposed in answer to this problem is
that of the self-organized critical ecosystem.

Self-organized criticality was first described by Bak, Tang, and
Wiesenfeld (1987), who studied the properties of a simple mathematical
model of avalanches in a pile of sand.  (It is from this work that we take
the name ``avalanches'' for the corresponding phenomenon in our evolving
system.)  In this model, grains of sand are deposited one by one on a
sand-pile, whose sides as a result grow steeper and steeper.  Eventually,
they are steep enough at some point on the pile that the addition of just
one more grain starts an avalanche at that point, and sand falls down the
slope.  As further grains are added, more and more such avalanches will
take place, small at first, but getting bigger as the pile gets steeper.
However, there is a limit to this process.  At some point---the so-called
critical point---the typical size of the avalanches becomes formally
infinite, which is to say there is bulk transport of sand down the pile.
This in turn reduces the slope of the pile, so that subsequent avalanches
will be smaller.  Then the process of building up the slope begins once
more.  As a result the sand pile can never pass the critical point at which
the avalanche size diverges; it organizes itself precisely to the point at
which the infinite avalanche takes place and the pile collapses, and then
stays close to that point ever afterwards.

It has been suggested that a similar process might be taking place in a
coevolving ecosystem.  Through mechanisms not yet well understood (although
there has been plenty of speculation) the ecosystem might be driven to
produce larger and larger coevolutionary avalanches, until it reaches a
critical point at which the typical avalanche size diverges.  At this point
a ``collapse'' takes place, analogous to the collapse of the sand in the
sand-pile, preventing any further change and holding the system close to
the critical point thereafter.  In other words, the system would organize
itself precisely to the point at which coevolutionary avalanches of
unlimited size take place, and these avalanches then might be responsible
for the widespread extinction seen in the fossil record.

We now review briefly a number of theoretical models which have been put
forward to explain how self-organization of the ecosystem might take place
in practice.

\subsection{The $NK$ model of Kauffman and Johnsen}
\label{nkmodel}
One of the first attempts to model large-scale coevolution quantitatively
is that of Kauffman and Johnsen~(1991), who created a model based on
Wright's picture of evolution on a rugged fitness landscape (Wright~1967,
1982).  In this model a fixed number of species evolve, each on its own
fitness landscape.  These landscapes are modeled explicitly using
techniques akin to those used in the study of spin glasses (Hertz and
Fischer~1991).  Each species possesses a certain number $N$ of genes, and
different fitnesses are assigned at random to different allelic states,
resulting in a fitness landscape in the genotype space.  The average
ruggedness of this landscape is controlled by a parameter $K$, which varies
the level of epistatic interactions between different genes.  In order to
produce coevolution, interactions between species are also introduced, each
species having the ability to affect the shape of the landscape of $S$
``neighboring'' species, through the interaction of $C$ of its genes with
$C$ of its neighbors'.

Under the presumed action of selection pressure, each species in the $NK$
model evolves by the sequential mutation of single genes to states of
higher fitness---it undergoes an ``adaptive walk''---and the ultimate
equilibrium state of the system is a Nash equilibrium in which each species
has reached a local fitness maximum and no single mutation will improve its
fitness any further.  Whether the system does in fact reach such a state
turns out to depend on the parameter $K$ which controls the ruggedness of
the landscapes.  (In this simplest version of the model, the value of $K$
is the same for all species, but versions have been studied which relax
this constraint.)

For high values of $K$ the landscapes in the $NK$ model are very rugged,
meaning that they possess many closely-spaced maxima and minima.  In this
situation it is relatively easy for most of the species simultaneously to
occupy local fitness maxima, at which point they stop evolving.  As a
result the system usually comes to rest after only a short coevolutionary
avalanche, and for this reason the high-$K$ regime is referred to as the
frozen regime---the system becomes frozen at a Nash equilibrium and stops
moving.

For lower values of $K$ the landscapes on which the individual species move
are smoother, which means that on average a species must evolve further
from its starting point to get to a fitness peak.  However, in so doing it
will change many of its genes, and is therefore likely to have an effect on
the fitness of other species.  As a result, coevolution becomes more common
as the value of $K$ gets lower, and coevolutionary avalanches get longer.
Eventually there comes a point at which coevolution never stops, and we
have an infinite avalanche.  At this point we have passed into the chaotic
regime.

A divergence of this kind in the avalanche size is of course precisely the
type of behavior we are looking for.  In the $NK$ model however it only
occurs at a certain critical value of the parameter $K$.  Is there any
reason to suppose that the ecosystem should be precisely at this critical
point?  There is in the $NK$ model no explicit self-organizing force which
pushes the system towards criticality, as there was in the sand-pile
described above, but Kauffman and Johnsen presented numerical results
which indicated that the fitness of the species in the ecosystem may be
maximized at the critical point, so that ordinary selection pressure would
drive them there.  In more recent work, Kauffman and Neumann~(1994) have
described a more complex version of the model in which in addition to
coevolution they introduced extinction by competitive replacement, and
their numerical experiments with this version seem to indicate that the
resulting distribution of the size $s$ of extinction events follows a
power law:
\begin{equation}
p(s) \propto s^{-\tau},
\label{powerlaw}
\end{equation}
where the exponent $\tau$ is about 1 in this case.  A power-law
distribution of event sizes is often a good indicator of critical behavior,
a point which will come up frequently in this paper.

\subsection{The self-organized critical model of Bak and Sneppen}
\label{bsmodel}
Another model of coevolutionary avalanche behavior which has attracted a
good deal of attention in the last few years is the model proposed by Bak
and Sneppen~(1993).  This model is related to the $NK$ model of the
previous section, but it incorporates one crucial new idea which, as it
turns out, is enough to cause the desired self-organization of the model
ecosystem to the critical point at which the mean avalanche size diverges. 

The new assumption of the Bak-Sneppen model is that the first species to
evolve, the one which starts the coevolutionary avalanche going in the
first place, is the species with the lowest fitness.  In the studies of
Kauffman and co-workers, by contrast, the first species to evolve was
chosen at random.  In addition, Bak and Sneppen made the assumption that
only the neighbors of that first species to evolve would be directly
affected by its evolution, limiting the immediate avalanche to only a
handful of species.  However, they then repeated this whole process,
starting another avalanche with the species with the next lowest fitness,
and so forth.  They observed that there was a greater probability of a
species having a low fitness if it had recently evolved, which means that
those species which took part in previous avalanches were more likely to be
chosen.  This gives rise to a sort of avalanche of avalanches, a wave of
evolution propagating across the ecosystem.  Although this is not exactly
the phenomenon which we were describing in Section~\ref{intro}, it is
possible that it could occur in nature, and that it could be responsible
for species extinction.  The elegant thing about the Bak-Sneppen model is
that it appears to be a true self-organized critical model in that,
regardless of the conditions it starts under, the system organizes itself
precisely to that state in which the mean avalanche size is infinite, and
the distribution of avalanche sizes follows a power law,
Equation~(\ref{powerlaw}), again with an exponent $\tau$ close to 1.

\subsection{Another self-organized critical model}
\label{newmanmodel}
Another possible mechanism for self-organized criticality in a coevolving
system has been suggested recently by Newman~(1997) and incorporated into a
model which makes very direct use of the coevolutionary avalanche idea,
although it does not employ the fitness landscape paradigm used by both of
the previous two models discussed.  In this model, the action of selection
pressure is assumed to favor a slow increase in the number of interactions
between species, with the result that the typical size of coevolutionary
avalanches (which depend on these interactions) grows over time.  However,
at the same time it is assumed that these coevolutionary avalanches cause
the extinction of a fraction of the species which they affect, though
mechanisms such as those described earlier.  When a species becomes extinct
in this way, all its interactions with other species vanish, and this
reduces the average size of coevolutionary avalanches once again.  The net
result is that the ecosystem drives itself just to the ``percolation
threshold'' at which the infinite avalanche takes place, and then stays
poised there.  This model also appears to be a true self-organized critical
model, capable of generating avalanches the size of the entire ecosystem.
The measured distribution of avalanche sizes follows a power-law with an
exponent of $\tau=\frac32$.

\subsection{The connection model of Sol\'e}
\label{solemodel}
Another model which focuses on the interactions between species has been
proposed by Sol\'e~\etal~(1996).  This model relies on a specific
assumption about the mechanism by which species become extinct: it is
assumed that species interactions can be both beneficial and harmful to a
species, and that if the harmful effects on a particular species of the
others around it outstrip the beneficial effects, the species will become
extinct.  The death of a species could be caused by, for example, its
inability to win sufficient resources in the face of overwhelming
competition, or its being hunted to extinction by an overzealous predator.
In detail the model is as follows.

A fixed number $N$ of species interact with one another in the model
ecosystem.  Each one interacts with a certain number $K$ of the others,
and the interactions may be harmful or beneficial.  Each interaction is
represented by a number whose magnitude is an indication of the strength
of the interaction, and which is either positive or negative depending on
whether the interaction is beneficial or harmful.  Note that there is no
need for the interactions to be symmetric, for the effect of species~A on
species~B to be the same as that of~B on~A.  To take an example, the
effect of a predator on its prey is clearly a harmful one, but the effect
of the prey on the predator is beneficial.

We want a species to become extinct if the harmful effects of other
species outweigh the beneficial ones, and this is achieved by a simple
rule.  If the sum of all the numbers representing the effects of other
species on any one species is less than zero, then that species becomes
extinct.  In order to keep the total number of species constant, the
extinct species is then replaced by speciation from one of the others.  In
order that the model doesn't grind to a halt when all possible extinctions
have occurred, it is also necessary to change species interactions
occasionally, by choosing one at random and giving it a new numerical
value.

This model is slightly different from the others we have considered, in
that the species interactions do not give rise to coevolution, but only to
extinction.  However, in simulations of the model, Sol\'e~\etal\ have found
that species tend to become extinct in waves, akin to the coevolutionary
avalanches we have been discussing, because the extinction of one species
removes its effect on any others with which it interacted, which will be
beneficial to some (if its previous presence was harmful) but harmful to
others.  As a result, some previously stable species will become extinct
following the extinction of one or more of their neighbors, and an
avalanche ensues.  Sol\'e and co-workers found that the mean size of these
avalanches diverges as the model comes to equilibrium, and the size
distribution follows a power law, with exponent $\tau\approx2$.  This again
may be evidence for critical behavior in the model, and an indication that
similar mechanisms might give rise to critical behavior in a real
ecosystem.

Recently, Manrubia and Paczuski~(1996) have proposed another model of
species interaction and extinction, which is, in essence, a simplified
version of the model of Sol\'e~\etal\ \ In their version the detailed
effect of the extinction of a species on all the neighboring species with
which it interacts is replaced by a random ``shock'' which makes all
species in the system more or less susceptible to extinction.  This version
of the model has the advantage of being analytically tractable.  It also
gives a power-law extinction size distribution, along with a number of
other interesting results.

It may appear that we have introduced a rather bewildering array of
different models here.  However, the important point is that, by the very
nature of critical phenomena, the predictions of all of these models are
somewhat similar.  All of them, for example, predict that the distribution
of the sizes of avalanches should follow a power law,
Equation~(\ref{powerlaw}), although they predict different values for the
exponent $\tau$.  Some of them also predict power-law distributions of
other quantities, such as the lifetimes of species.  So, without even
knowing which if any of the many models is a good representation of
processes taking place in the real ecosystem, it is possible to examine
fossil and other data for these tell-tale signs of critical behavior.  In
this paper we outline some of the evidence which has been put forward in
favor of self-organized critical behavior in terrestrial evolution, showing
that a number of the relevant quantities do indeed possess power-law
distributions.  However, we believe that it is not justified to conclude
from this evidence that evolution is a critical process.  To demonstrate
this, we propose a new and very simple model of evolution and extinction in
which species die out as a result of environmental stresses and not because
of coevolution or any other endogenous effects.  Our model does not in fact
incorporate any elements which mimic the effects of coevolution in the
ecosystem, but nonetheless it reproduces accurately all the evidence which
has been claimed to be the result of self-organized critical behavior.  As
a result, we conclude, there is no reason to invoke self-organized
criticality as an explanation for the observed data.

\section{Evidence in favor of self-organized criticality}
\label{evidence}
The one result on which all of the models described above concur is that
the distribution of the sizes of avalanches should follow a power law of
the form given in Equation~(\ref{powerlaw}).  Unfortunately, it has not
proved possible to observe coevolutionary avalanches directly in any but a
small number of cases, so we do not have any statistical data on what the
distribution of their sizes might be in nature.  What we can study is the
distribution of the sizes of extinction events, for which we have
moderately good fossil data.  Some of the models described in the last
section, such as those of Kauffman and Neumann, Sol\'e~\etal, and Manrubia
and Paczuski, clearly predict that the extinction distribution should also
follow a power law, and this is a prediction which we can test.  Others do
not make an explicit connection to extinction, but nonetheless indicate
that a power-law extinction distribution might be expected, without making
a precise prediction about its exponent.  It therefore makes sense to ask
whether the distribution of the sizes of extinction events in the fossil
record does indeed follow a power law.

\begin{figure}
\begin{center}
\psfigure{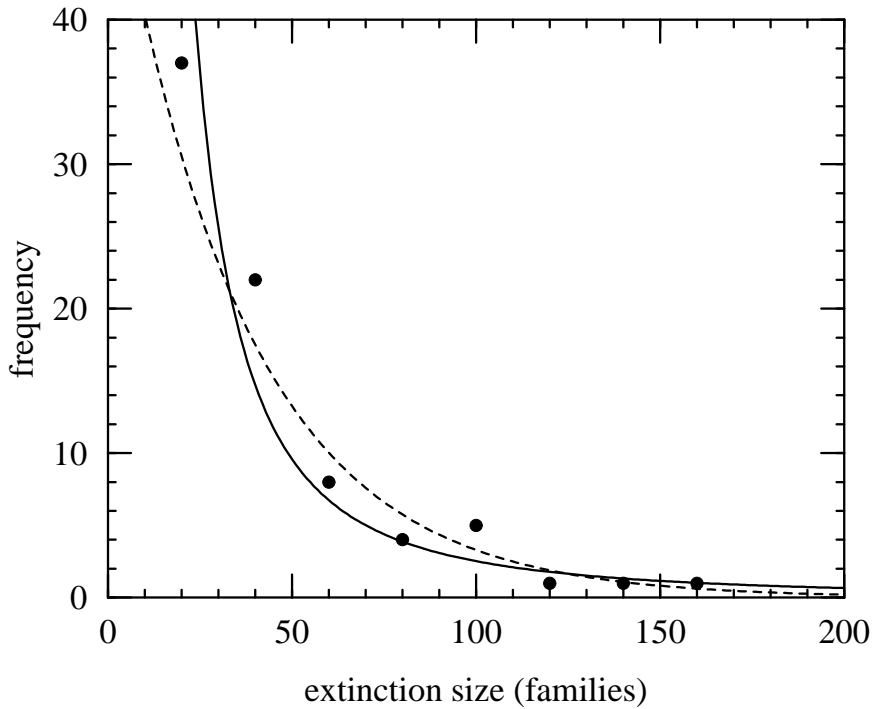}{11.5cm}
\end{center}
\caption{Frequency distribution of extinction rates over 79 geologic stages
  during the phanerozoic, with the best fitting power-law (solid line) and
  exponential (dashed line) curves.  After Sol\'e and Bascompte~(1996).
\label{sbfit}}
\end{figure}

\begin{figure}
\begin{center}
\psfigure{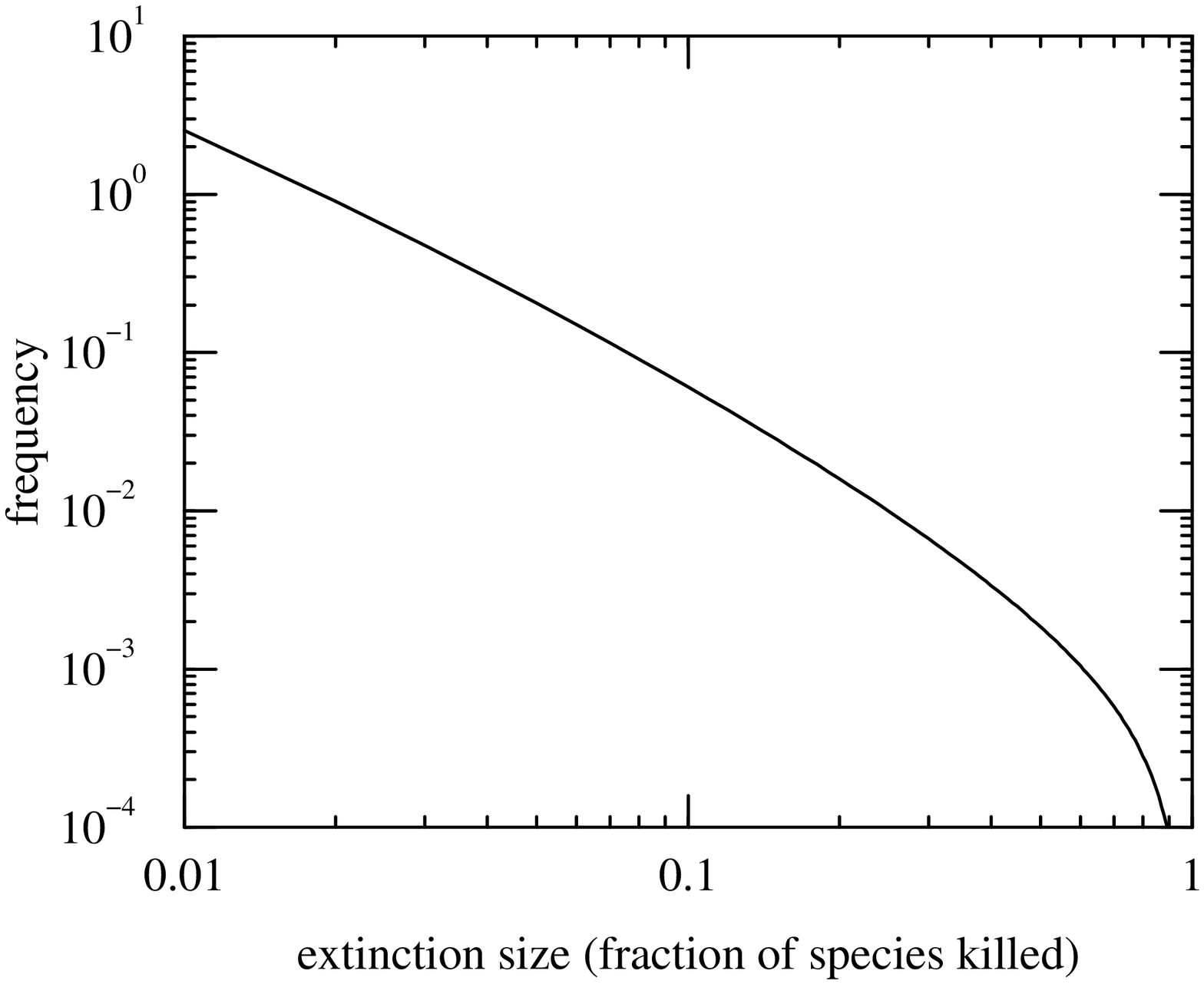}{11.5cm}
\end{center}
\caption{The extinction distribution corresponding to the kill curve
  extracted from Sepkoski's fossil data by Raup~(1991).  The curve is
  approximately power-law in form with an exponent of $\tau=1.9\pm0.4$.
\label{raupfit}}
\end{figure}

Data on the extinction of paleozoic and mesozoic marine invertebrates
compiled by Sepkoski~(1993) and analysed by Raup~(1986), has been used by
Sol\'e and Bascompte~(1996) to show that the fossil extinction distribution
is compatible with a power law, and that the exponent $\tau$ is equal to
about $1.95$.  Their fit to the data is reproduced in Figure~\ref{sbfit}.
They also point out however, that, given the rather poor statistical
quality of the data, the distribution is also fitted acceptably by an
exponential distribution, which is definitely incompatible with
self-organized critical theories.  Raup~(1991) has used the same data to
construct a so-called ``kill curve'', which is a cumulative frequency
distribution of extinctions, and Newman~(1996) has shown that the
extinction size distribution can be deduced from this kill curve by a
simple mathematical transformation.  It turns out that Raup's curve is
approximately equivalent to a power-law distribution of extinction events
with an exponent of $\tau=1.9\pm0.4$---see Figure~\ref{raupfit}.  In the
same paper, the author also made use of a Monte Carlo technique to fit a
power-law form to the fossil data and extracted a figure $\tau=2.0\pm0.2$
for the best fit.  Thus it is probably fair to say that the distribution of
extinction events in the fossil record is compatible with a power-law form,
although the data are not good enough to rule out other possible functional
forms.

The figure $\tau\approx2$ is interesting because it makes quantitative
comparison possible between models of extinction and empirical data.  As it
turns out, most of the models discussed above are compatible with this
figure.  The models proposed by Bak and Sneppen~(1993) and by Newman~(1997)
are purely models of evolution and make no numerical prediction about the
distribution of extinction events.  The models of Sol\'e~\etal~(1996) and
of Manrubia and Paczuski~(1996) both predict values of $\tau$ close to two
in good agreement with the fossil data.  The only model discussed here
which is ruled out by the data is that of Kauffman and Neumann~(1994),
which predicts that $\tau$ should take a value close to one.

\begin{figure}
\begin{center}
\psfigure{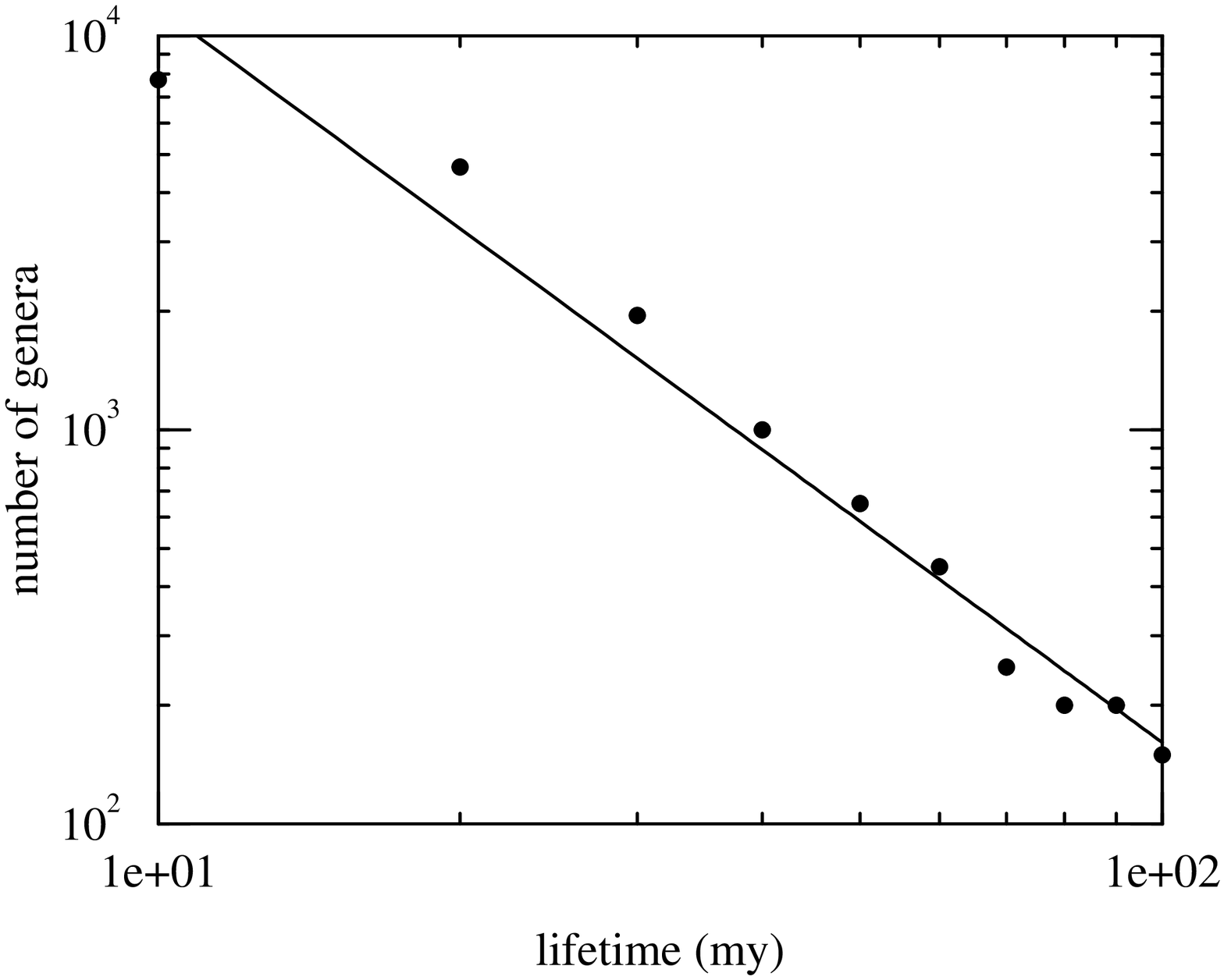}{11.5cm}
\end{center}
\caption{The distribution of genus lifetimes drawn from Sepkoski's data.
  The distribution is approximately power-law in form with an exponent of
  $\alpha=1.9\pm0.1$ (solid line).  After Sneppen~\etal~(1995).
\label{sbfjfit}}
\end{figure}

Another form of evidence comes from the distribution of the lifetimes of
taxa in the fossil record.  A number of the models described in the last
section make the prediction that the lifetimes of species should also have
a power-law distribution, and this too can be tested by examining fossil
data.  In order to measure the lifetime of a taxon accurately, one needs a
reasonably generous sample of fossil representatives.  (Poorly represented
taxa are susceptible to the Signor-Lipps effect~(Signor and Lipps~1982)
which tends to result in underestimated lifetimes.)  As a result, it is
common to work with higher taxa, usually genera or families, when making
lifetime estimates, rather than species.  Again using Sepkoski's data,
Sneppen~\etal~(1995) have examined the distribution of genus lifetimes over
the entire phanerozoic, and have concluded that the distribution is
approximately power-law in form, with an exponent $\alpha$ measured to be
in the vicinity of two.  The data are reproduced in Figure~\ref{sbfjfit}.

\begin{figure}
\begin{center}
\psfigure{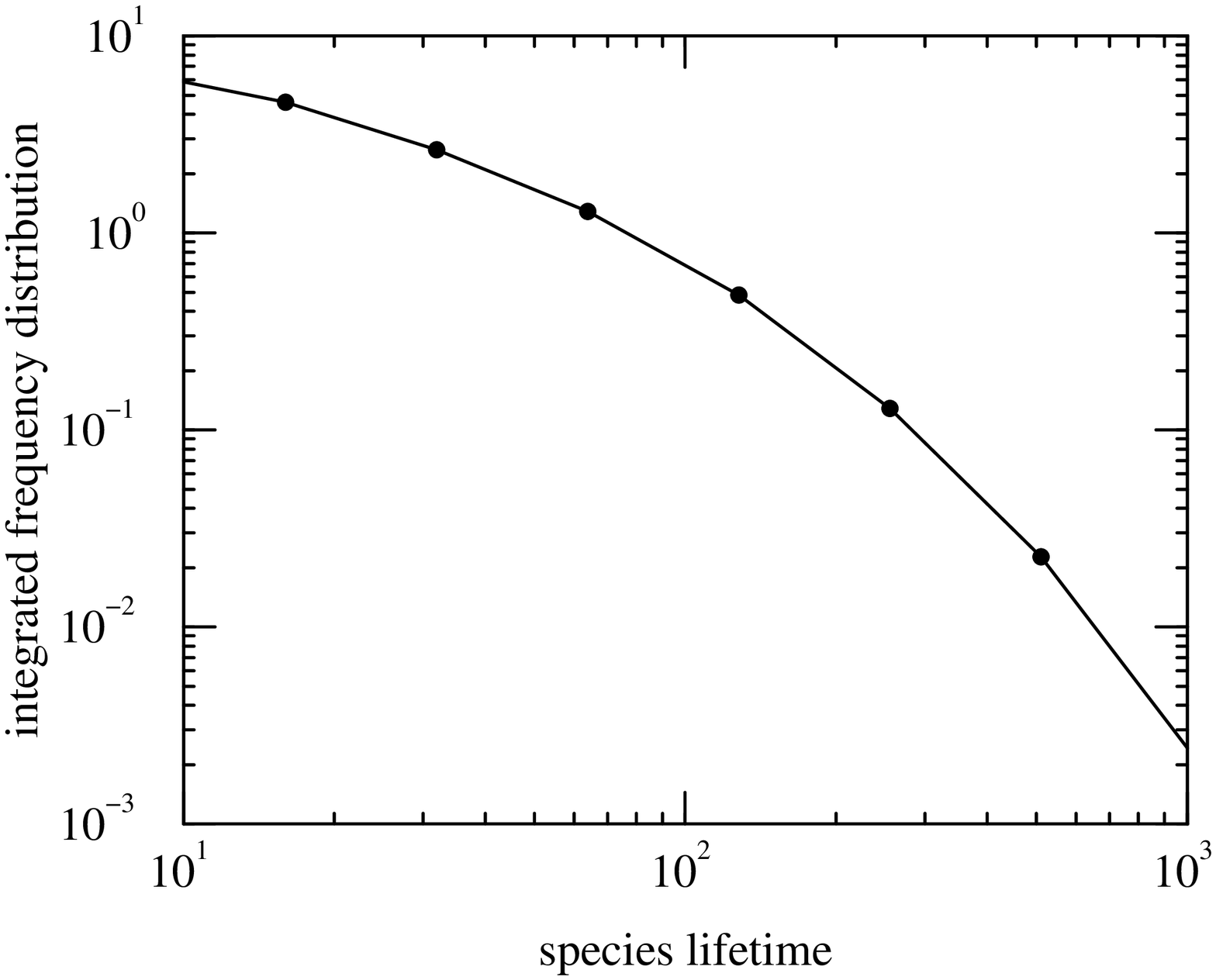}{11.5cm}
\end{center}
\caption{The integrated distribution of species lifetimes in simulations
  performed using the Tierra artificial life system.  The distribution is
  approximately power-law in form with an exponent $\alpha$ near one.  The
  fall-off in the curve for long lifetimes in caused by finite-time effects
  in the simulations.  After Adami~(1995).
\label{adamifit}}
\end{figure}

Another example of a system in which species have a power-law distribution
of lifetimes has been observed recently by Adami~(1995), not with
biological data, but with data on the evolution of competing computer
programs in the Tierra artificial life environment created by Ray~(1994).
In these simulations, self-reproducing programs compete for limited
resources in the form of CPU time and memory space on a computer, and those
which reproduce most successfully rapidly dominate the system.  In the
course of a number of very large Tierra simulations, Adami observed the
lifetimes of the dominant species in the system and demonstrated that a
histogram of these lifetimes approximately follows a power law with an
exponent $\alpha$ near one---see Figure~\ref{adamifit}.  Although these
data come from a very different kind of ecosystem to the biological ones
which are our principal concern here, many of the same considerations apply
to the two cases and it is possible that results from one can shed light on
the other.  We should point out however, that, intriguing though Adami's
power-law forms are, it is not clear whether they are the result of
self-organized critical behavior (Newman~\etal~1997).

\begin{figure}
\begin{center}
\psfigure{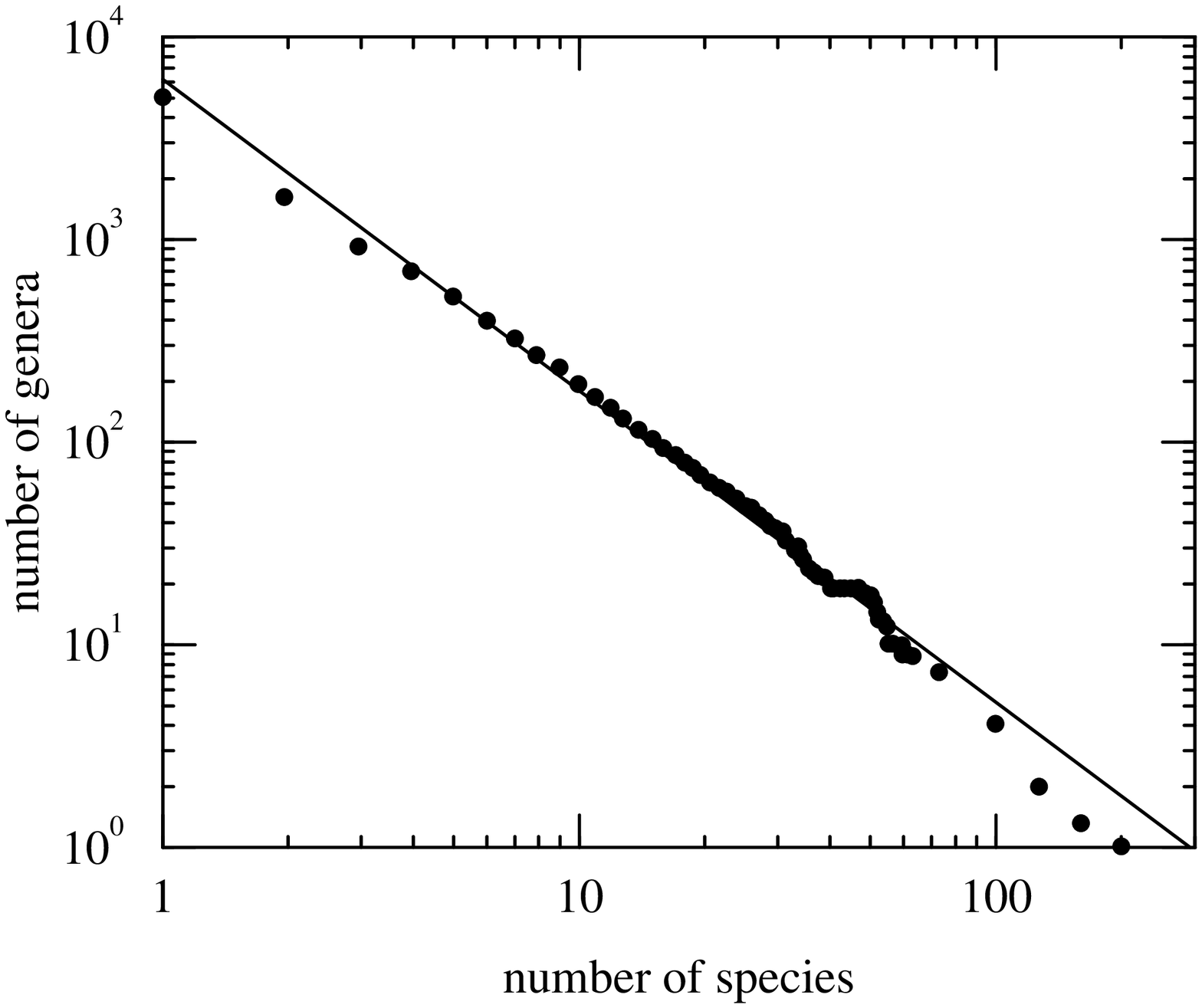}{11.5cm}
\end{center}
\caption{Histogram of the number of species per genus for flowering plants.
  The distribution is approximately power-law in form with a measured
  exponent of $1.5\pm0.1$.  After Willis~(1922).
\label{wyfit}}
\end{figure}

A third type of evidence for critical behavior, independent of the fossil
data on which the previous two rested, comes from taxonomic analyses.
Using data from living biota, Willis~(1922) noted that the distribution of
the number genera with a certain number of species follows a power law with
an exponent whose value is variable, but typically around $\frac32$
(Figure~\ref{wyfit}).  In combination with the empirical law of age and
area proposed by Willis, which in one of its forms states that the number
of species in a genus increases linearly with the age of the genus, this
result again implies a power-law distribution in the lifetimes of genera.
Note however that this result should be considered independent of the
direct measurement of the genus lifetime distribution shown in
Figure~\ref{sbfjfit}, since it is derived from entirely different data.
Burlando~(1990, 1993) has demonstrated that the power-law distribution of
species within genera extends also to higher taxa, indicating a fractal
self-similarity in the taxonomic hierarchy which may possibly also be the
result of critical behavior in evolution.

Recently Sol\'e~\etal~(1997) have also studied the power spectrum of the
time series data for phanerozoic extinction events.  They conclude that the
spectrum approximately follows a $1/f$ law, and tentatively propose that
this may indicate critical behavior.  As they point out, $1/f$ noise is a
widespread phenomenon, occurring in many systems which are not critical,
but nonetheless their results add one more data point to the argument.

In addition to these quantitative kinds of evidence, a number of authors
have pointed to general trends in the evolutionary record which may be
indicators of critical behavior.  Chief amongst these are the punctuated
equilibria highlighted in the work of Gould and Eldredge~(1993), which
consist of bursts of evolutionary activity separated by periods of
comparatively little change.  Self-organized critical models typically show
intermittent patterns of activity which are somewhat similar, and it has
been suggested (see, for example, Bak and Sneppen~(1993)) that the two
phenomena are in fact one and the same.  It should be pointed out that
traditional evolutionary theory is not especially in need of an explanation
of punctuated equilibria: the standard picture of evolution on a rugged
fitness landscape implies that species will spend long periods of time
close to particular fitness peaks before making a rapid movement to a new
peak.  Whether this movement is provoked by coevolutionary pressures or not
makes little difference, the punctuation will be present either way.
However, it is certainly possible that some of the punctuations visible in
the paleontological record are the result of intermittent coevolutionary
avalanches, and hence a sign of critical behavior.

One might think then, given these different types of evidence, that there
was moderately good cause to believe that biological evolution does indeed
drive the ecosystem to a critical point, resulting in power-law
distributions of various quantities.  However, as we mentioned above, we do
not believe this to be a justified conclusion.  In support of this view, we
now introduce and study in some detail a simple model of evolution and
extinction which reproduces all of the evidence above although it is not a
self-organized critical model and does not contain any element which mimics
the proposed coevolutionary avalanche behavior.  A brief account of this
model has appeared previously (Newman~1996).

\section{A model for evolution and extinction}
\label{model}
The model we propose is a simple one.  Its assumptions are few in number
and straightforward.  To begin with, we assume that the ultimate cause of
extinction for any species is environmental stress of one kind or another,
a very conventional point of view (see, for example, Hoffmann and
Parsons~(1991)).  Stresses of a variety of different kinds have been
associated with most of the major extinction events in the Earth's history
(Jablonski~1986).  They include climate change, changes in sea level,
bolide impact and a variety of other factors.  There is no reason why
coevolutionary avalanches should not also be a contributing factor to
extinction.  It is certainly possible, as discussed above, that a large
coevolutionary avalanche could place a strain on the ecosystem and cause
the extinction of a number of species.  This possibility is not excluded
from our model, although neither is it given any special treatment.  As far
as the model is concerned, stresses are stresses.  In fact, the only
feature distinguishing one stress from another within our model is their
strength, which if the model is to have any realism at all, must presumably
vary with time.  Sometimes the climate will be particularly harsh and at
other times it will be clement.  Sometimes there will be large rocks
raining down from space while at others only small ones, or none at all.
In the simplest version of our model, all of these effects are represented
by just one quantity, $\eta(t)$, which measures the level of stress at time
$t$.

We introduce a number $N$ of species into the model, all of which feel the
same stresses, represented by $\eta(t)$.  Any species will become extinct
if hit by a sufficiently large stress.  However, we assume that the
threshold level of stress required to drive a species extinct varies from
one species to another.  For the $i^{\rm th}$ species we denote this
threshold by $x_i$.  If at any time the stress level $\eta(t)$ exceeds the
extinction threshold for a particular species, then that species becomes
extinct.  Since it is observed that the number of species a habitat can
support is roughly a constant over time (Benton~1995), we replace these
extinct species with equal numbers of new ones, which are assumed to have
speciated from survivors.  Thus the number of species remains constant at
$N$.

This is essentially all there is to our model---extinction as a result of
stresses placed on the system, and replacement by speciation.  However,
there are a number of blanks which still need to be filled in.  First, how
is the value of $\eta(t)$ at any particular time chosen, and what values
do the threshold variables $x_i$ take?  We divide time in the usual
fashion into discrete time-steps, and, since we have no reason to do
otherwise, choose the stress level $\eta$ to be a new random number at
each step.  Assuming that small stresses are more common than large ones,
we draw these random numbers from a distribution $p_{\rm stress}(\eta)$
which falls off away from zero, though it is not necessary that the
distribution be strictly monotonic.  The exact form of $p_{\rm
stress}(\eta)$ does not, as we will see, matter as far as the principal
predictions of the model are concerned, however some plausible forms for
the function might be a Poissonian distribution, or Gaussian white noise.

The threshold variables $x_i$ are chosen initially at random.  When new
species appear in the aftermath of an extinction event they need to be
assigned values of $x_i$, and there are a couple of reasonable ways in
which we might do it.  One way would be to have them inherit values from
other surviving species, from which they are assumed to have speciated.
Another way might be simply to assign new values drawn at random from some
distribution $p_{\rm thresh}(x)$.  For example, values might be chosen to
lie uniformly in the interval between zero and one.  We have experimented
with threshold values chosen according to both of these methods.  To a
large extent we again find that the predictions of the model do not depend
on the choice we make.

There is one further element which we need to add to our model in order to
make it work.  As we have described it, the species in the system start off
with randomly chosen thresholds $x_i$ and, through the extinction mechanism
described above, those with the lowest thresholds are systematically
removed from the population and replaced by new ones.  As a result, the
number of species with low thresholds for extinction decreases over time
and so the size of the extinction events taking place dwindles.
Ultimately, extinctions will cease altogether, a behavior which we know not
to be representative of a real ecosystem.  The solution to this problem
comes, we believe, from evolution.  In a real ecosystem, extinction as a
result of applied stress certainly can increase the mean fitness with
respect to stress, as we see here in our model, but we can also assume that
in the intervals between large stress events species will evolve under
other selection pressures, possibly at the expense of their ability to
withstand stress.  In other words, the necessary business of adapting to
the environment can, as a side effect, change a species' ability to survive
the next large stress placed on it by its environment, and that change,
although it could be for the better, could also be for the worse.  There
are a couple of possible ways to represent this situation in the model.
One, which we can think of as the gradualist viewpoint, is to have the
values of all the variables $x_i$ wander slowly over time, by adding or
subtracting a small random amount to each one at each time-step.  Another
possibility, the punctuationalist viewpoint, would hold most species
constant at any given time-step, but allow a small fraction $f$ to evolve
to new forms with completely different values of $x_i$.  We experiment with
both these possibilities in the next section, and again demonstrate that,
to a large extent, the model's predictions are independent of the choice we
make.

This then completes our model.  In one of its simplest variations, the
model could be summarized as follows.  We take $N$ species, labeled by
$i=1\ldots N$ and initially assign to each a threshold for stress $x_i$
chosen at random from a distribution $p_{\rm thresh}(x)$.
\begin{enumerate}
\item At each time step we choose a number $\eta$ at random from a
  distribution $p_{\rm stress}(\eta)$ to represent the stress level at that
  time, and all species possessing thresholds for extinction $x_i$ below
  that level become extinct.  The fraction $s$ of the total $N$ species
  which become extinct in this way is the size of the extinction event
  occurring in this time-step.  The extinct species are replaced with new
  ones whose thresholds for extinction $x_i$ are chosen at random from the
  distribution $p_{\rm thresh}(x)$ again.
\item A small fraction $f$ of the species, also chosen at random, evolve to
  new forms and are assigned new values of $x_i$ chosen at random from the
  distribution $p_{\rm thresh}(x)$.
\end{enumerate}

This version of the model is in fact identical to the model used by Newman
and Sneppen~(1996, Sneppen and Newman~1997) to study the dynamics of
earthquakes.  In these papers we investigated the properties of the model
analytically in some detail.  Rather than reproduce that discussion, the
reader is referred for details to those papers.  Here we investigate
instead the model's properties as they apply to the issue of biological
extinction.

\section{Properties and predictions of the model}
\label{results}
The first and most important feature of the model which we should point
out is that it is not a self-organized critical model.  The model does not
show coevolutionary avalanches of the kind which it is argued are
responsible for self-organizing behavior, and indeed the species in the
model do not interact with one another at all.  Each species develops
entirely independently of all the others and its ultimate fate will the
same regardless of what any of the others do.  This is, of course, an
oversimplification of the true situation.  There is no doubt that real
species do interact and do coevolve.  In Section~\ref{variations} we will
examine a more sophisticated variation of our model which reintroduces
species interaction.  However, the simple version presented here serves a
very useful purpose, since as we will see, even without interaction
between the species it reproduces all the forms of evidence for
self-organized criticality put forward in Section~\ref{evidence},
indicating that the mechanisms present here---stress-driven extinction,
repopulation, and random uncorrelated evolution---are on their own
perfectly adequate to explain the data.

\begin{figure}
\begin{center}
\psfigure{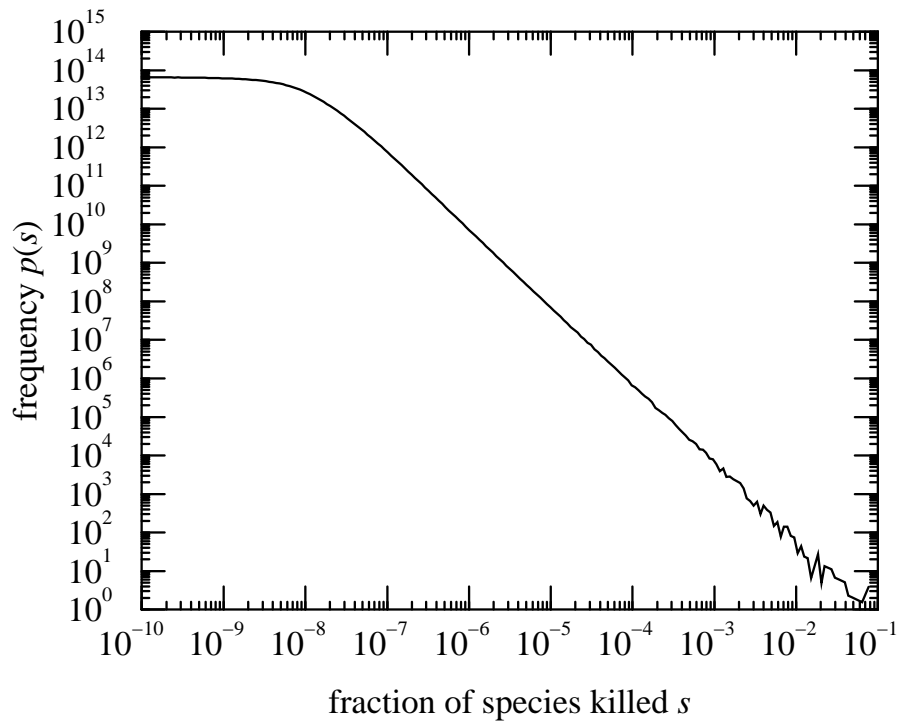}{11.5cm}
\end{center}
\caption{The distribution of the sizes of extinction events during a
  simulation of the model described in Section~\protect\ref{model}.  The
  distribution follows a power law closely over many decades, before
  flattening out around $s=10^{-8}$.  In this example, in which the
  stresses on the system were drawn from a normal distribution, the power
  law has an exponent of $2.02\pm0.02$.
\label{sdist1}}
\end{figure}

\subsection{Distribution of extinction sizes}
\label{sizes}
The fundamental prediction of our model is that a certain number of species
may be expected to become extinct at each time-step, and that the fraction
$s$ which does so depends on the level of stress placed on the system
during that step, and on the number of species present in the population
whose ability to withstand that stress is low.  In Figure~\ref{sdist1} we
show a histogram of the sizes of the extinction events taking place over
the course of a computer simulation of the model lasting ten million
time-steps.  The histogram is plotted on logarithmic scales, and the
straight line form of the graph indicates that the histogram follows a
power law of the form given in Equation~(\ref{powerlaw}).  The only
deviation is for very small extinction sizes, in this case below about one
species in $10^8$, for which the distribution becomes flat.  However,
extinctions this small are well below the noise level in our fossil data,
so to the resolution of the data the prediction of our model is that the
distribution of extinction sizes should be power-law in form.  The
distribution of applied stresses for the simulation shown in
Figure~\ref{sdist1} was normal with standard deviation $\sigma=0.1$ and
mean zero:
\begin{equation}
p_{\rm stress}(\eta) \propto \exp\biggl[ -{\eta^2\over2\sigma^2} \biggr].
\label{normal}
\end{equation}
The exponent $\tau$ of the power-law distribution in Figure~\ref{sdist1}
can be measured with some accuracy and is found to be $2.02\pm0.02$.
Recall from Section~\ref{evidence} that the distribution of the sizes of
extinction events in the fossil record has been found to be compatible with
a power law form, and that this has been taken by some as an indication of
self-organized critical behavior.  Here, however, we see the same result
emerging from a non-critical model of the extinction process, and
furthermore, the measured value of $\tau$ is in excellent agreement with
the value of $2.0\pm0.2$ extracted from the fossil data.

\begin{figure}
\begin{center}
\psfigure{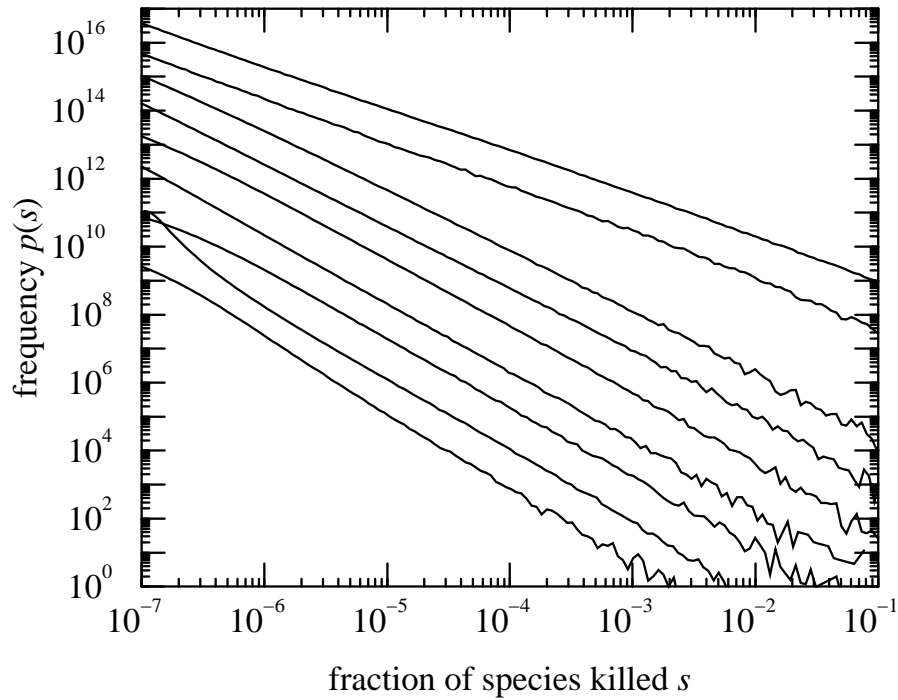}{11.5cm}
\end{center}
\caption{The distribution of the sizes of extinction events in the model
  for a variety of different types of applied stress distributions.  The
  distributions used include normal centered around zero, normal centered
  away from zero, Poissonian, exponential, stretched exponential, and
  Lorentzian.
\label{sdist2}}
\end{figure}

In Figure~\ref{sdist2}, we show the distribution of extinction sizes for a
wide variety of other stress distributions $p_{\rm stress}(\eta)$.  As the
figure makes clear, the power-law distribution of extinction sizes is a
ubiquitous phenomenon, and does not rely on the presence of any particular
stress distribution.  Furthermore, although the exponent of the power law
function varies somewhat as the stress distribution is changed, it is
always quite close to two, in agreement with the value observed in the
fossil data.  In a previous paper (Sneppen and Newman~1997) we have given
an analytical explanation of this property of the model, as well as
simulation results for the distribution of extinction sizes in the model
for a variety of different choices of the fraction $f$ of species which
evolve at each time step.  As we show, the power-law form of the extinction
distribution is present in all cases, with exponent in the vicinity of two,
except when $f$ becomes very large (comparable to one).

\begin{figure}
\begin{center}
\psfigure{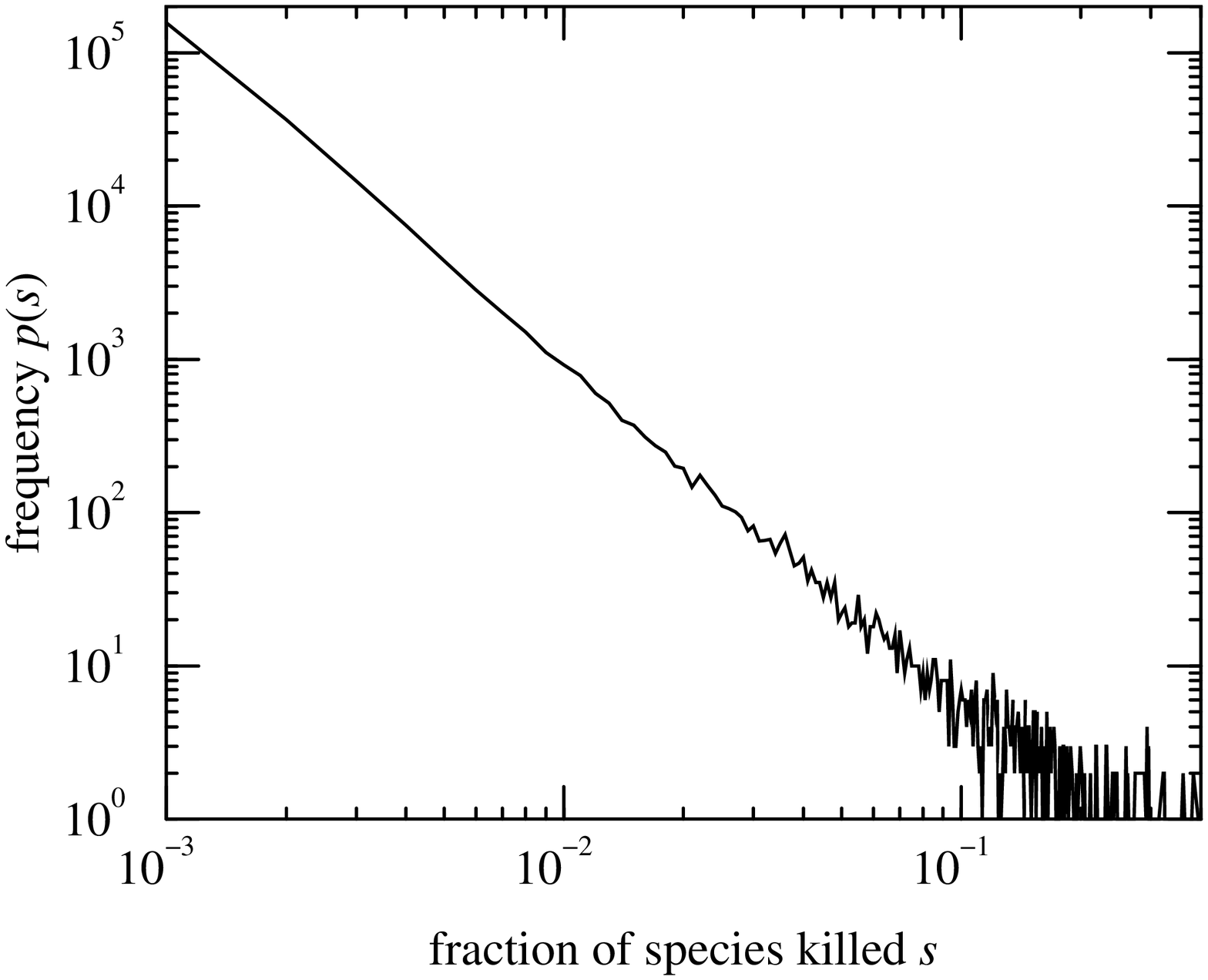}{11.5cm}
\end{center}
\caption{The distribution of the sizes of extinction events for the
  variation of the model in which newly appearing species inherit their
  threshold values from survivors of the last extinction event.
\label{sdist3}}
\end{figure}

\begin{figure}
\begin{center}
\psfigure{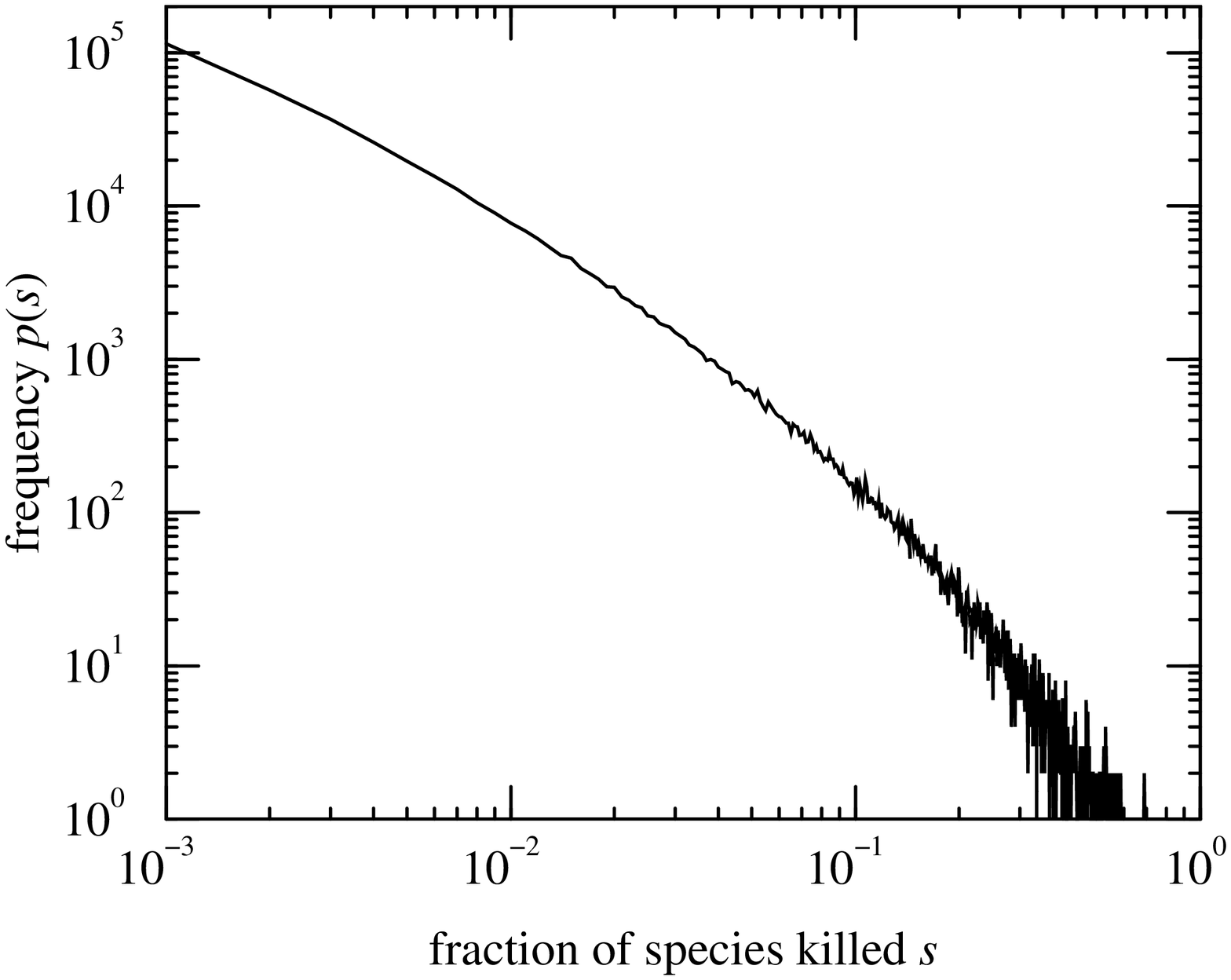}{11.5cm}
\end{center}
\caption{The distribution of the sizes of extinction events for the
  variation of the model in which evolution takes place in a gradual
  fashion, the values of $x$ for each species performing a slow random
  walk, rather than changing abruptly as in most of our other simulations.
\label{sdist4}}
\end{figure}

In Figures~\ref{sdist3} and~\ref{sdist4} we show distributions of
extinction sizes drawn from simulations of the model in which newly
appearing species inherit values of $x_i$ from the survivors of the last
extinction event, or in which evolution takes place by the gradualist
process described in Section~\ref{model} where the thresholds of all
species perform a slow random walk as time progresses.  As the figures
show, the power-law form of the extinction distribution is robust against
all of these variations in the dynamics of the model.

\begin{figure}
\begin{center}
\psfigure{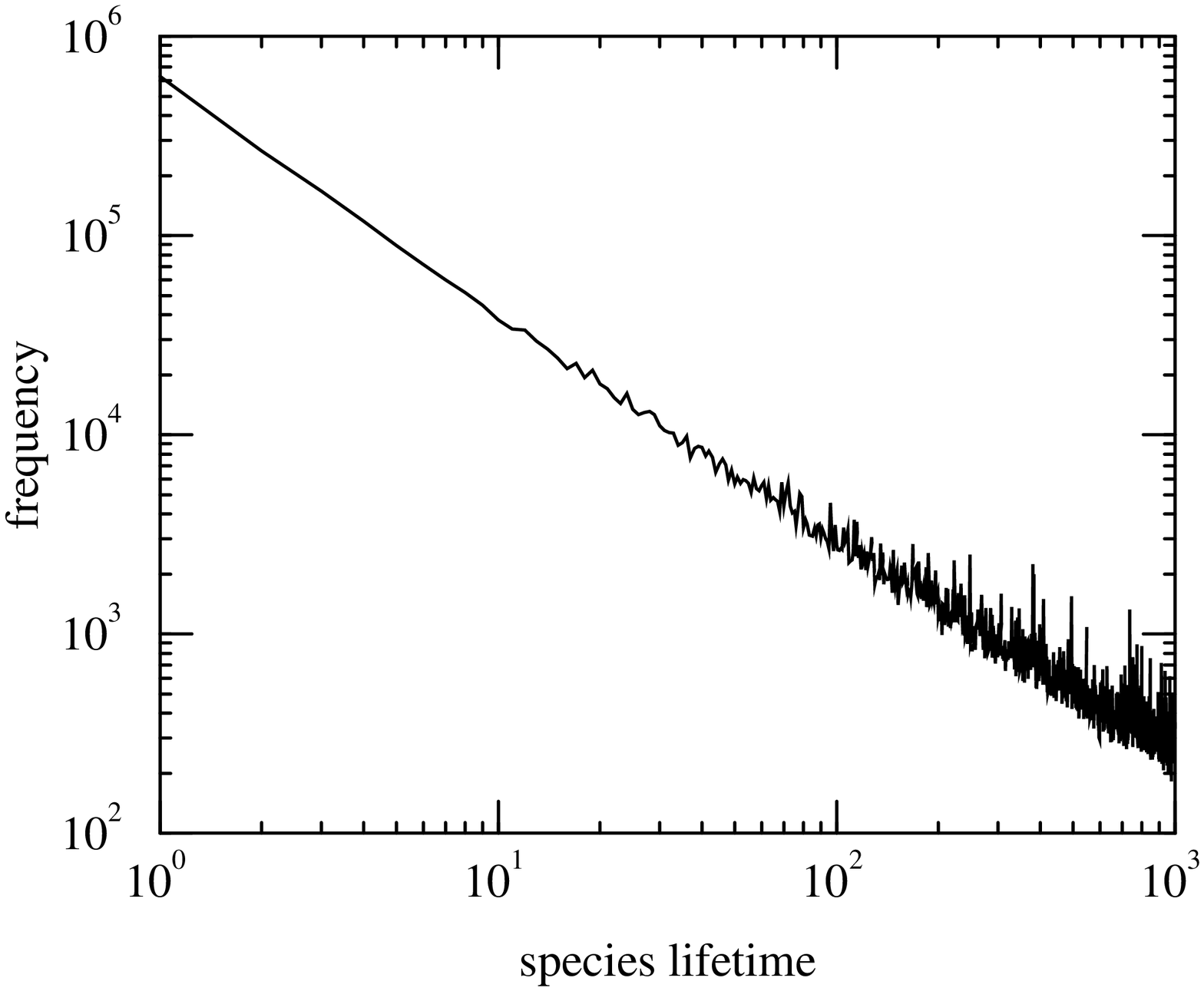}{11.5cm}
\end{center}
\caption{The distribution of species lifetimes measured in simulations of
  the model.  The distribution is power-law in form, with an exponent of
  $1.03\pm0.05$.
\label{ldist}}
\end{figure}

\subsection{Species lifetimes}
\label{slife}
It is also a straightforward matter to measure the lifetimes of species in
our model.  Counting the number of time-steps between the first
introduction of a species and its eventual extinction, we have constructed
a histogram, Figure~\ref{ldist}, of species lifetimes.  Again the axes are
logarithmic, and the straight-line form indicates that the distribution
follows a power law.  The exponent $\alpha$ of this power law is measured
to be $1.03\pm0.05$, which is, for example, close to the distribution of
lifetimes measured by Adami~(1995) in his work on artificial life.  As
discussed in Section~\ref{evidence}, measurement of species lifetimes in
the fossil data is prone to error and studies have tended to concentrate
more on the higher taxa.  In the next section we consider how information
on genera, including the distribution of genus lifetimes, can be extracted
from our model.

\subsection{Genera}
\label{genera}
The model as we have described it contains no information about taxonomic
structure.  However it is not difficult to extend it so that it does.  We
start off by assigning every species to its own unique genus, and
thereafter when a new species appears it is assumed to have speciated from
one of the previously existing ones, and therefore it should share the same
genus as that parent species.  As before, we make the simplest assumption
and choose the parent species at random from the available possibilities.
This on its own results in an ever dwindling number of genera, since genera
can become extinct if all their member species vanish, but new ones can
never appear.  In reality this doesn't happen because every once in a while
a species appears which is declared to be the founding member of a new
genus.  This process can be emulated in the model by choosing a small
fraction $g$ of new species at random to found genera.  (Choosing them at
random may seem rather an extreme route to take, but on the other hand it
may not be so very different from the behavior of a real taxonomist.)  The
result is a model in which genera appear, flourish, and become extinct,
just as species do.

\begin{figure}
\begin{center}
\psfigure{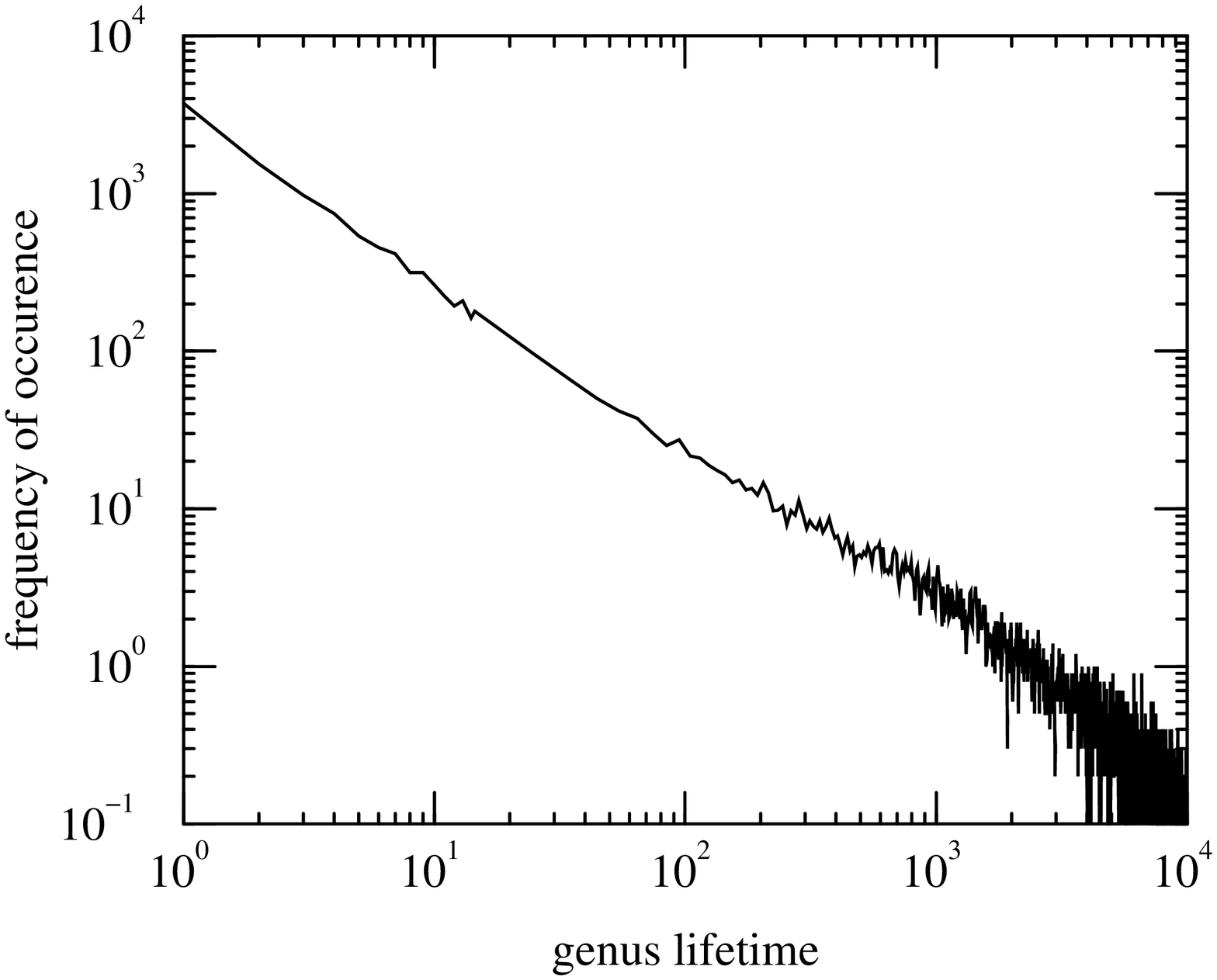}{11.5cm}
\end{center}
\caption{The distribution of genus lifetimes measured in simulations of
  the model.  The distribution is power-law in form, with an exponent of
  $1.0\pm0.1$.
\label{glife}}
\end{figure}

In Figure~\ref{glife} we show the distribution of the lifetimes of genera
drawn from a simulation of the model, plotted again on logarithmic scales.
As with the lifetimes of species the distribution follows a power law.
The exponent in this case is measured to be $1.0\pm0.1$.

\begin{figure}
\begin{center}
\psfigure{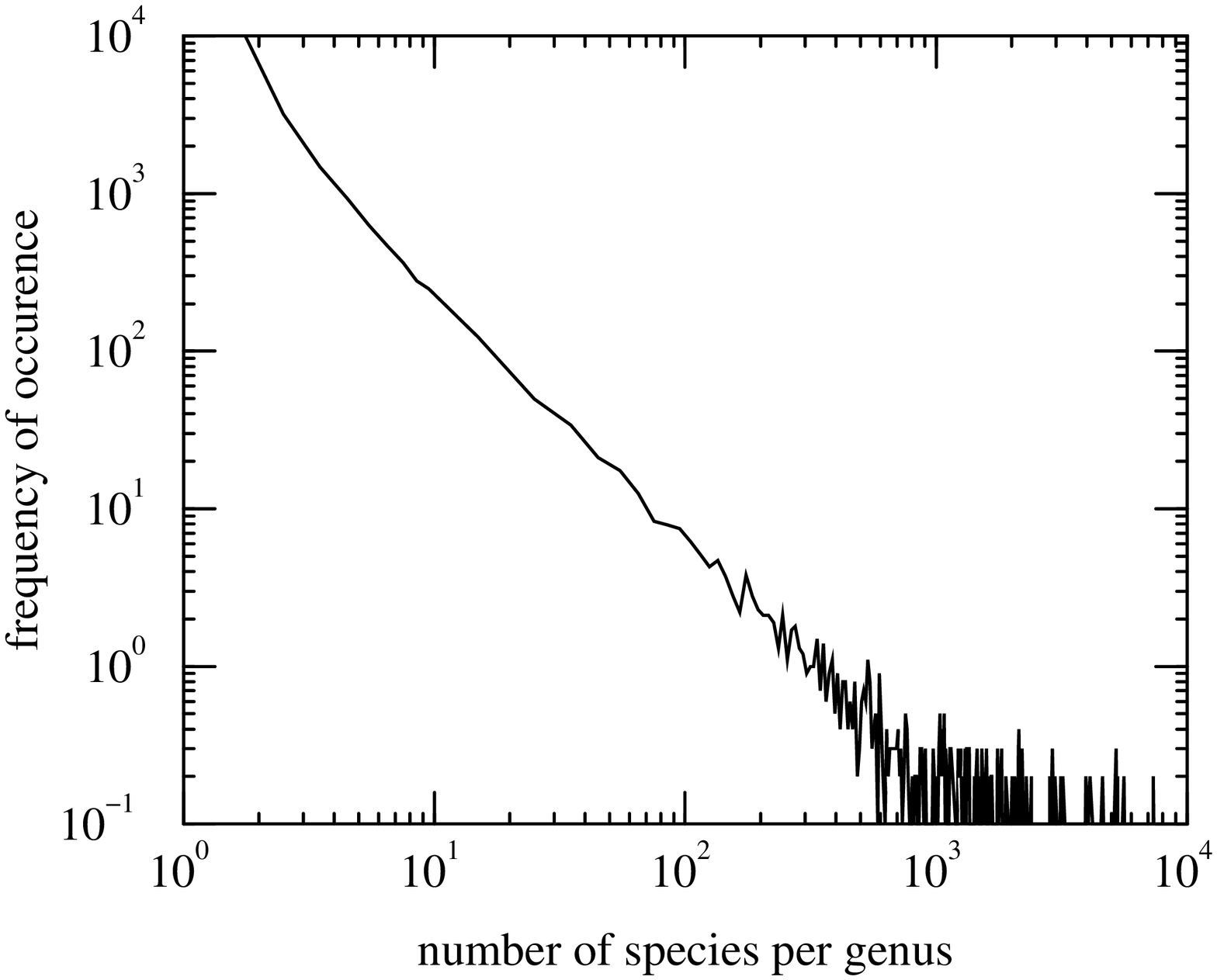}{11.5cm}
\end{center}
\caption{Histogram of the number of species per genus in a simulation of
  our model.  The distribution has an exponent of $1.6\pm0.1$.
\label{gpers}}
\end{figure}

In Figure~\ref{gpers} we show a histogram of the numbers of species in each
genus in the same simulation.  This too follows a power law, with an
exponent measured in this case to be $1.6\pm0.1$.  This result is in
agreement with the studies of modern taxonomic trees performed by
Willis~(1922) and more recently by Burlando~(1990, 1993) (see
Section~\ref{evidence}), which showed that this distribution does indeed
follow a power law, with a measured exponent in the vicinity of $\frac32$.

We see that our simple model of evolution and extinction agrees both
qualitatively and quantitatively with the various forms of evidence put
forward in Section~\ref{evidence} even though it is not a self-organized
critical model.  However, the model also makes some independent predictions
about extinction which may help to determine whether the processes which it
models actually do take place in the real world.  One of the most striking
of these predictions concerns the existence of ``aftershock extinctions''.

\subsection{Aftershock extinctions}
\label{aftershocks}
In the model we have proposed, stresses on the system render extinct those
species which are not strong enough to survive them, in effect selecting
for those which are.  However, even species which are well able to
withstand stress may lose that ability, because in the periods when the
selection pressure of the stress is absent they may evolve in other ways,
and this can make them more susceptible to stress.  The result is that the
longer the time between one large stress on the system and the next, the
more species will have taken advantage of the lull to exert themselves
adaptively in other directions, reducing their tolerance for stress and
making them more likely to become extinct next time around.  As a result,
long periods in which the stress level is low tend to be followed by large
extinction events.  Unfortunately, this is not an effect which is likely to
be easily observed, since it is very hard to know what level of stress
species were feeling given only their fossils and the accompanying
geological record.  However, there is another related effect which may be
visible.

\begin{figure}
\begin{center}
\psfigure{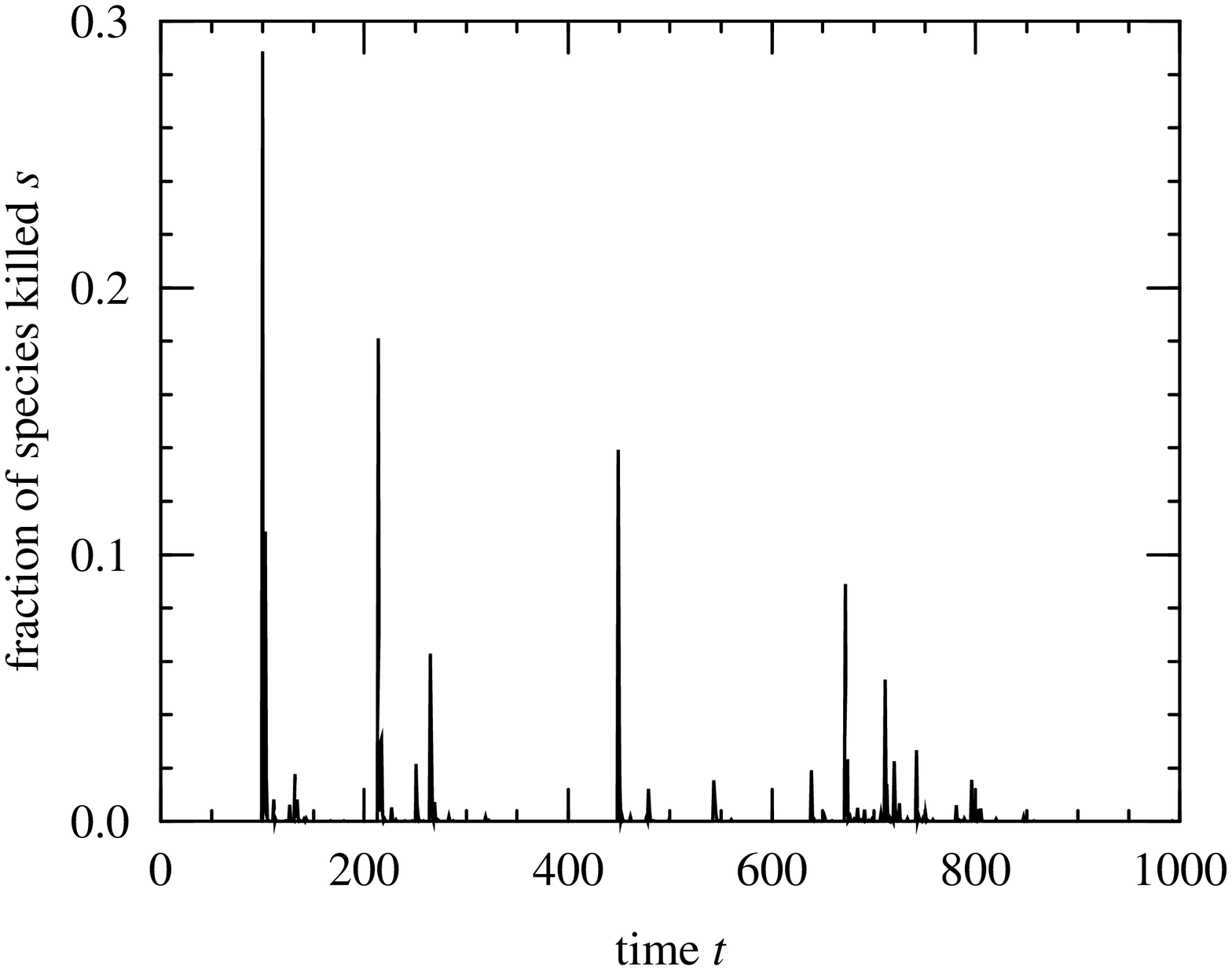}{11.5cm}
\end{center}
\caption{A section of the time-series of extinction events in a
  simulation of the model.  The aftershocks following the first large event
  are clearly visible.  Notice also that some of the aftershocks themselves
  generate a smaller series of after-aftershocks.
\label{aftershock1}}
\end{figure}

When a large extinction event does take place, it opens the way for a large
number of new species to appear, a phenomenon which can be seen clearly in
the fossil record.  However, it is possible that some of the species which
appear to fill newly-vacated niches may not be very well adapted to the
lives which they are trying to lead, having not had very long, in
evolutionary terms, to adapt to them.  In particular, these opportunistic
species have not yet been subject in their short lives to any dramatic
environmental stresses, and although some of them may, fortuitously, be
well-adapted to survive such stresses, others may not be, with the result
that they will get wiped out when the next stress of even moderate size
appears on the horizon.  Thus we expect that in the aftermath of a large
extinction event there will appear opportunists which last only a brief
time before disappearing themselves in another, smaller extinction event.
This is what we call an aftershock extinction, and the effect is clearly
visible in our model.  In Figure~\ref{aftershock1} we show an example of a
series of aftershocks drawn from one of our simulations.  In this
particular example it is also clear that the aftershocks themselves give
rise to after-aftershocks, and so forth in a decaying series.  It is
possible that aftershock extinctions might also be visible in the fossil
record.  To our knowledge no one has looked for such an effect, but it
might make an interesting study.

\begin{figure}
\begin{center}
\psfigure{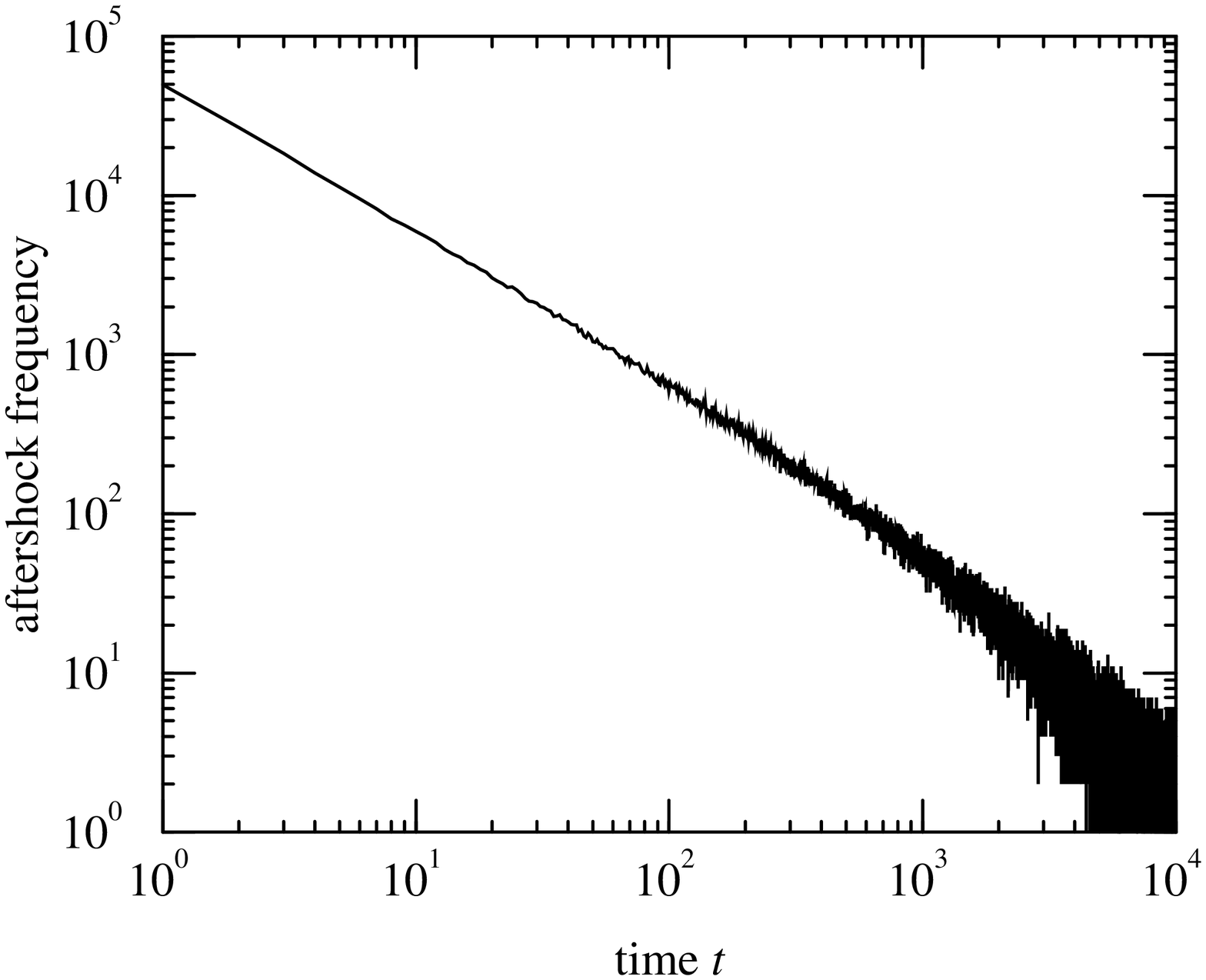}{11.5cm}
\end{center}
\caption{A histogram of the time distribution of aftershocks following a
  large event.  The distribution follows a power law with an exponent
  very close to one.
\label{aftershock2}}
\end{figure}

The time spacing of the aftershocks is of interest too.  We have measured
the time in between large extinction events and each of the smaller
aftershocks which come after them.  Figure~\ref{aftershock2} is a histogram
of these times, and again it follows a power law.  The exponent in this
case is $-1$, which is to say that the probability per unit time of the
occurrence of an aftershock extinction following a large extinction event
goes down as $t^{-1}$ with time after the initial large event.  In a
previous paper (Newman and Sneppen~1996) we have given an argument
explaining why we believe this power-law to be exact, with exponent $-1$
regardless of the distribution of stress levels, or any of the other
parameters of the model.  It would be very interesting if it were possible
to observe this behavior in the fossil record too, although it seems
unlikely that the resolution of the currently available data is up to this
task (Raup, private communication).

\section{Variations on the model}
\label{variations}
The model we have studied in the previous sections of this paper was about
as simple as we could make it, and deliberately so, since our primary aim
has been to show that the data which others have used in favor of
self-organized critical theories of evolution can been explained by much
simpler assumptions.  However, there are crucial features of the real
ecosystem which are missing from our model, and it is important to find out
whether these have any effect on the behavior predicted by our model.  Of
course, the real ecosystem is arbitrarily complicated and there is no way
we can ever remotely approach its complexity with a model such as the one
described here; it is the hope of modeling work such as ours that the gross
features of the extinction process are dominated by a few basic mechanisms
and that the other details of the way in which individual species evolve
make only a small contribution to the overall picture.  Nonetheless, there
are undoubtedly some very important factors which are missing from the
model as it stands, and it would be good to demonstrate that these do
indeed not affect our fundamental predictions.  In this section we examine
briefly two such factors, both of which lead to generalized versions of the
model.  The first is species interactions.

In Section~\ref{intro} we discussed the importance of interspecies
interactions in producing coevolution.  The self-organized critical
theories of ecosystem organization rest upon the contention that these
interactions are the dominant force shaping the biosphere.  Is it not
possible then that the presence of such interactions could make the
behavior of real ecosystems entirely different from that of our model?  In
order to address this question we have introduced interspecies interactions
into the model in a way akin to that suggested by Bak and Sneppen~(1993).
The model is now placed on a lattice.  It could be a single line, or a
square grid, or a random lattice.  It turns out to make little difference.
The dynamics of the model is as before except that now, as well as wiping
out all those species with thresholds $x_i$ for extinction which are less
than the stress level $\eta$, we also wipe out their neighbors on the
lattice.  The rationale behind this move is that when a species becomes
extinct there exists the possibility that it will take with it some of the
others which depend on it; the extinction of a particular plant species for
example might result in the extinction of the insect which lays its eggs on
the leaves.  (The much less dramatic reality of course is that the
extinction of one species usually just forces minor adaptations in others.
However, we are exaggerating the effect here in order to investigate its
influence on our model.)

\begin{figure}
\begin{center}
\psfigure{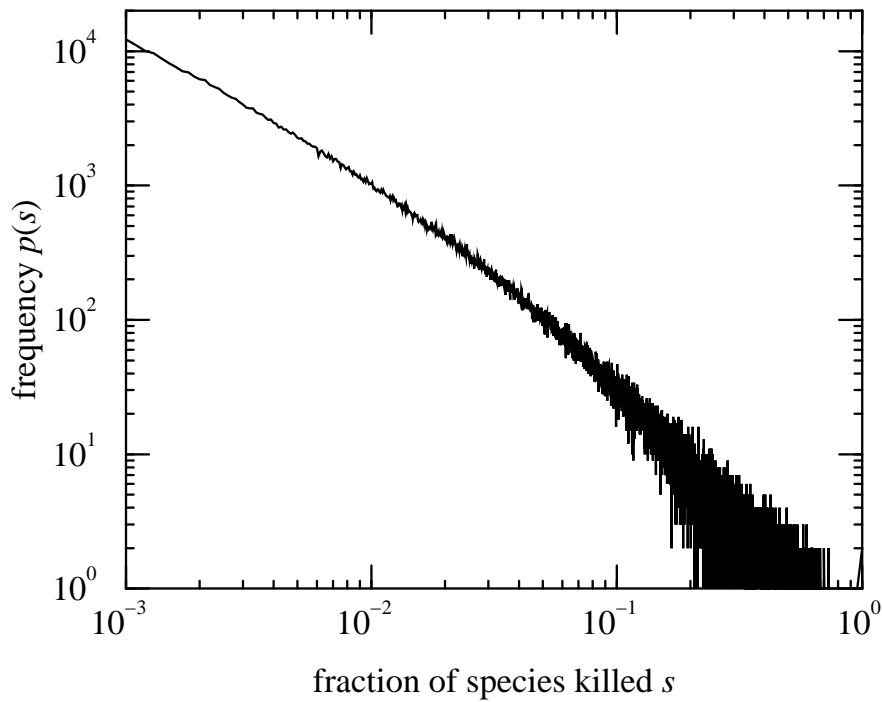}{11.5cm}
\end{center}
\caption{The distribution of the sizes of extinction events for a
  variation of the model in which the species are placed on a lattice and
  the extinction of one species as a result of the applied stress gives
  rise to the extinction of all the neighboring species of that one.  As
  the figure shows, the distribution of event sizes still follows a power
  law.
\label{var1}}
\end{figure}

In Figure~\ref{var1} we show the distribution of extinction sizes
calculated in a simulation of this variation of the model.  As the figure
makes clear, our basic prediction of a power-law distribution of
extinction sizes is unchanged.  The exponent is still in the vicinity of
two for any choice of stress distribution $p_{\rm stress}(\eta)$, in
agreement with the fossil data.  The addition of species interactions
does of course have some effect on the model.  In particular there is now
a correlation between the species which become extinct: if a species has,
say, four neighbors with which it interacts, and all of them become
extinct when it does, then there will be a group of five species which all
became extinct at once.  It is possible that such groups, arising through
this `knock-on' extinction effect, might be observed empirically.
(Deforestation, which might be viewed as a form of extinction, is well
known to have substantial knock-on effects, for example.)  However, the
large-scale predictions of our model, and their agreement with fossil and
other data, are unaltered.

Another obvious problem with our model is that it regards all stresses as
being equivalent, whereas in reality this is clearly not the case.  In real
life the stresses on an ecosystem are of many different types and different
species will have different tolerances for each type.  A species living in
the warm shallow waters at the edge of the ocean may be devastated by a
three meter drop in sea level, while another living above the snow-line at
three thousand meters may not feel a thing.  The meteor which lands in
central Africa may spell disaster for those close to the impact, but others
living in Siberia may be indifferent.  In order to incorporate this concept
in our model, we turn our single stress level $\eta$ into many levels
$\eta_1$, $\eta_2$, etc.\ each one representing the level of a different
type of stress, and each one chosen independently at random at each time
step.  One such level might represent stress arising from changes in sea
level for example, and another changes in climate, and so forth.  Each
species also has many threshold variables, which we can denote $x_{i,1}$,
$x_{i,2}$ and so on, measuring the species' tolerance for the corresponding
type of stress.  Now if at any time the level of any one type of stress
exceeds a species' tolerance for it, then the species becomes extinct.

\begin{figure}
\begin{center}
\psfigure{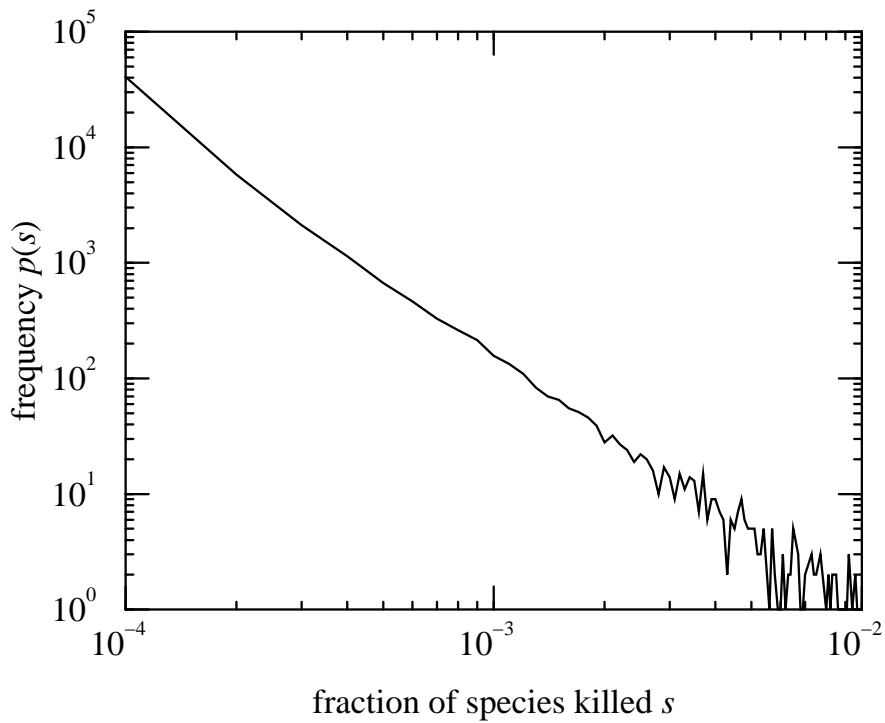}{11.5cm}
\end{center}
\caption{The distribution of the sizes of extinction events for a
  variation of the model in which there are many different kinds of stress,
  and a species may become extinct as a result of any one of them.  In this
  particular simulation there were ten different kinds of stress.
\label{var2}}
\end{figure}

Figure~\ref{var2} shows an example of the distribution of extinction sizes
generated in a simulation of this version of the model with, in this case,
ten different types of stress.  Again we see that the power-law form of
the distribution is preserved, and the exponent is still close to two.  In
fact it seems not unreasonable that this should be the case.  Presumably
the type of stress for which the threshold is lowest is the one which is
most likely to drive a particular species extinct, and if we make the
approximation of simply ignoring all the different types of stress except
this one then, mathematically speaking, the model becomes identical to the
simple form in which there is only one type of stress.  Crude though this
approximation is, it gives an indication that the behavior of the two
versions should indeed be similar.

\section{Conclusions}
\label{concs}
We have reviewed the arguments and evidence which have been put forward in
favor of self-organized critical processes in evolution.  They revolve
primarily around the demonstration of the existence of power-law
distributions in a variety of quantities, including the sizes of extinction
events seen in the fossil record, the lifetimes of fossil genera, and the
number of species per genus in taxonomic trees.  We have then introduced a
new and simple model in which extinction is caused by random stresses
placed on the ecosystem by its environment.  This model is not
self-organized critical, and indeed does not in its simplest form contain
any interactions between species whatsoever.  Nonetheless, as we have
demonstrated, it reproduces all of the above evidence well.  We therefore
suggest that this evidence should not be taken (as it has been by some) to
indicate critical behavior in terrestrial evolution.

\section*{Acknowledgements}
The author would like to thank David Raup, Kim Sneppen, and Ricard Sol\'e
for interesting discussions, and Chris Adami and Simon Fraser for supplying
the data used to produce Figures~\ref{adamifit} and~\ref{wyfit}.  This work
was supported in part by the Cornell Theory Center, by the NSF under grant
number ASC-9404936, and by the Santa Fe Institute and DARPA under grant
number ONR N00014--95--1--0975.

\vspace{0.5in}

\section*{References}
\smallskip

\begin{list}{}{\leftmargin=2em \itemindent=-\leftmargin%
\itemsep=4pt \parsep=0pt}

\item {\frenchspacing Adami, C. 1995 Self-organized criticality in living
    systems. {\it Phys. Lett. A\/} {\bf203}, 29.}

\item {\frenchspacing Alvarez, L. W., Alvarez, W., Asara, F., and
    Michel, H. V. 1980 Extraterrestrial cause for the
    Cretaceous-Tertiary extinction.  {\it Science} {\bf 208},
    1095--1108.}
  
\item {\frenchspacing Bak, P. and Paczuski, M. 1996 Mass extinctions vs.
    uniformitarianism in biological evolution.  To appear in {\it Physics
      of Biological Systems,} Springer-Verlag, Heidelberg.}

\item {\frenchspacing Bak, P. and Sneppen, K. 1993 Punctuated equilibrium
and criticality in a simple model of evolution.  {\it Phys. Rev. Lett.}
{\bf71}, 4083.}

\item {\frenchspacing Bak, P., Tang, C., and Wiesenfeld, K. 1987
    Self-Organized criticality: An explanation of $1/f$ noise.  {\it
      Phys. Rev.  Lett.} {\bf59}, 381.}
  
\item {\frenchspacing Benton, M. J. 1995 Diversification and extinction in
    the history of life.  {\it Science\/} {\bf268}, 52.}
  
\item {\frenchspacing Burlando, B. 1990 The fractal dimension of taxonomic
    systems. {\it J. Theor. Biol.} {\bf146}, 99.}

\item {\frenchspacing Burlando, B. 1993 The fractal geometry of
    evolution. {\it J. Theor. Biol.} {\bf163}, 161.}

\item {\frenchspacing Fischer, K. H. and Hertz, J. A. 1991 {\it Spin
      Glasses}, Cambridge University Press, Cambridge.}

\item {\frenchspacing Glen, W. 1994 What the
    impact/volcanism/mass-extinction debates are about.  In {\it The
      Mass Extinction Debates,} Glen, W., (ed.), Stanford University
    Press, Stanford.}
  
\item {\frenchspacing Gould, S. J. and Eldredge, N. 1993 Punctuated
    equilibrium comes of age. {\it Nature\/} {\bf366}, 223.}
  
\item {\frenchspacing Hoffmann, A. A. and Parsons, P. A. 1991 {\it
      Evolutionary Genetics and Environmental Stress,} Oxford University
    Press, Oxford.}

\item {\frenchspacing Jablonski, D. 1986 Causes and consequences of mass
    extinctions.  In {\it Dynamics of extinction,} Elliott, D. K. (ed.),
    Wiley, New York.}

\item {\frenchspacing Kauffman, S. A. 1992 {\it The Origins of Order,}
Oxford University Press, Oxford.}

\item {\frenchspacing Kauffman, S. A. and Johnsen, S. 1991 Coevolution
    to the edge of chaos: Coupled fitness landscapes, poised states,
    and coevolutionary avalanches.  {\it J. Theor. Biol.} {\bf149},
    467.}

\item {\frenchspacing Kauffman, S. A. and Neumann, K. 1994 Unpublished
    results.}

\item {\frenchspacing Manrubia, S. C. and Paczuski, M. 1996 A simple model
    of large scale organization in evolution.  Submitted to {\it
    Phys. Rev. Lett.}.}

\item {\frenchspacing Newman, M. E. J. 1996 Self-organized criticality,
    evolution and the fossil extinction record.  {\it Proc. R. Soc. Lond.
    B\/} {\bf263}, 1605.}

\item {\frenchspacing Newman, M. E. J. 1997 Self-organized criticality
    in evolution.  Santa Fe Institute working paper.}

\item {\frenchspacing Newman, M. E. J., Fraser, S. M., Sneppen, K., and
    Tozier, W. A. 1997 Comment on ``Self-organized criticality in living
    systems'' by C. Adami.  {\it Phys. Lett. A,} in press.}

\item {\frenchspacing Newman, M. E. J. and Sneppen, K. 1996 Avalanches,
    scaling, and coherent noise.  {\it Phys. Rev. E} {\bf54}, 6226.}

\item {\frenchspacing Raup, D. M. 1986 Biological extinction in Earth
history.  {\it Science\/} {\bf231}, 1528.}

\item {\frenchspacing Raup, D. M. 1991 A kill curve for Phanerozoic
marine species.  {\it Paleobiology\/} {\bf17}, 37.}

\item {\frenchspacing Ray, T. S. 1994 An evolutionary approach to synthetic
    biology. {\it Artificial Life\/} {\bf1}, 195.}
  
\item {\frenchspacing Sepkoski, J. J. 1993 Ten years in the library: New
    data confirm paleontological patterns.  {\it Paleobiology\/} {\bf19},
    43.}

\item {\frenchspacing Sharpton, V. L., Dalrymple, G. B., Martin, L.
    E., Ryder, G., Schuraytz, B. C., and Urrutia-Fucugauchi, J. 1992
    New links between the Chicxulub impact structure and the
    Cretaceous/Tertiary boundary.  {\it Nature} {\bf 359}, 819.}
  
\item {\frenchspacing Signor, P. W. and Lipps, J. H. 1982 Sampling bias,
    gradual extinction patterns, and catastrophes in the fossil record.
    In {\it Geological Implications of Impacts of Large Asteroids and
      Comets on the Earth,} Silver, L. T. and Schultz, P. H. (eds.),
    Geological Society of America Special Paper {\bf 190}, 291.}
  
\item {\frenchspacing Sneppen, K., Bak, P., Flyvbjerg, H., and Jansen, M. H.
    1995 Evolution as a self-organized critical phenomenon.  {\it
    Proc. Nat. Acad. Sci.} {\bf92}, 5209.}

\item {\frenchspacing Sneppen, K. and Newman, M. E. J. 1997 Coherent noise,
    scale invariance and intermittency in large systems.  {\it Physica D,}
    in press.}

\item {\frenchspacing Sol\'e, R. V., Manrubia, S. C., Benton, M., and Bak,
    P. 1997 Private communication.}

\item {\frenchspacing Sol\'e, R. V. and Bascompte, J. 1996 Are critical
phenomena relevant to large-scale evolution?  {\it Proc. Roy. Soc. B\/}
{\bf263}, 161.}

\item {\frenchspacing Sol\'e, R. V., Bascompte, J., and Manrubia, S. C. 1996
    Extinction: bad genes or weak chaos?  {\it Proc. Roy. Soc. B\/}
    {\bf263}, 1407.}

\item {\frenchspacing Willis, J. C. 1922 {\it Age and Area,} Cambridge
    University Press, Cambridge.}

\item {\frenchspacing Wright, S. 1967  Surfaces of selective value.  {\it
Proc. Nat. Acad. Sci.} {\bf58}, 165.}

\item {\frenchspacing Wright, S. 1982  Character change, speciation and the
higher taxa.  {\it Evolution\/} {\bf36}, 427.}

\end{list}

\end{document}